\documentclass[a4paper]{article} 
\pdfoutput=1

\usepackage{times}
\usepackage{graphicx}
\usepackage[table]{xcolor}
\usepackage{makecell}
\usepackage{soul}
\usepackage{xcolor}
\usepackage{multirow}
\usepackage{amsmath}
\usepackage{amssymb}

\title{Analysing PolSAR data from vegetation\\by using the subaperture decomposition approach}
\author{J. David Ballester-Berman
\\ \vspace{0.3cm}\small{Universitat d'Alacant, Spain}\\
\small{\today} }
\date{}

\definecolor{red_light}{rgb}{0.88,0.5,0.5}
\definecolor{grey}{rgb}{0.55,0.55,0.55}

\begin{document}

\maketitle

\pagestyle{headings}
\setcounter{page}{1}
\pagenumbering{arabic}

\begin{abstract}
A common assumption in radar remote sensing studies for vegetation is that radar returns originate from a target made up by a set of uniformly distributed isotropic scatterers. Nonetheless, several studies in the literature have noted that orientation effects and heterogeneities have a noticeable impact in backscattering signatures according to the specific vegetation type and sensor frequency. In this paper we have employed the subaperture decomposition technique (i.e. a time-frequency analysis) and the 3-D Barakat degree of polarisation to assess the variation of the volume backscatterig power as a function of the azimuth look angle. Three different datasets, i.e. multi-frequency indoor acquisitions over short vegetation samples, and P-band airborne data and L-band satellite data over boreal and tropical forest, respectively, have been employed in this study. We have argued that despite depolarising effects may be only sensed through
a small portion of the synthetic aperture, they can lead to overestimated retrievals of the volume scattering for the full
resolution image. This has direct implications in the existing model-based and model-free polarimetric SAR decompositions.
%
\end{abstract}

\section{Introduction and Motivation}


Characterisation of vegetation covers by means of active radar remote sensing relies on assumptions aimed at reducing the problem complexity while achieving a reasonable accuracy in the retrieval of the physical properties of the observed target. One of the most commonly used assumptions in vegetated scenes is to consider the radar signature of the imaged pixel being originated by a homogeneously distributed isotropic target. However, a number of works in the literature have reported that orientation effects are not uncommon in natural covers and must be considered depending on the particular vegetation type and the sensor frequency \cite{ar:treuhaft_cloude}.

These anisotropic effects in distributed targets were revealed in \cite{ar:ferro-famil2003} by means of the subaperture decomposition approach (equivalent to a time-frequency analysis). Authors proposed a combination of subaperture decomposition  and the H-$\alpha$ decomposition to analyse the different polarimetric behaviour of natural media as the azimuth look angle changes. It was shown that Bragg resonances in radar signatures can be detected in agricultural fields due to nonstationary polarimetric behaviour by jointly observing the total backscattered power and the entropy and $\alpha$ values. From these analysis, a nonstationarity test for distributed targets was also developed and validated in \cite{ar:ferro-famil2003}. It is noted, however, that this type of spectral analysis was first employed in an early work \cite{pro:arnaud} for ship detection through the complex interferometric coherence computed between the radar acquisitions from the front and rear parts of the synthetic aperture. The reader is referred to \cite{ar:marino2015} for a comprehensive review on the use of spectral analysis for ship detection.

Simultaneously to \cite{ar:ferro-famil2003}, in \cite{ar:sourys2003} the sublook processing was used to design an algorithm, based on the complex correlation between sublooks, to discriminate point targets from natural covers where fully developed speckle dominates the radar response. Similarly, the detection of coherent scatterers was addressed in \cite{ar:zandona06} by means of the sublook coherence and employing the entropy of the sublook images as a measure to assess the spectral correlation between sublooks. It is noted, however, that this approach only considered the spectral analysis in the range dimension.

Anisotropic radar signatures in forest covers were also revealed at P-band in \cite{ar:garestier08_p} where evidences were provided regarding the need to consider these effects for PolInSAR-based forest height inversion. Similarly, in \cite{ar:frey} P-band SAR tomograms over forests showed that radar backscattering was dominated by ground returns where coherent backscattering occur. According to \cite{ar:dalessandro2013}, these complex scattering phenomena at P-band are a consequence of local forest density variations, suggesting the need to consider a varying wave extinction along the horizontal direction. In the same vein, other studies \cite{ar:lopez_rvog_test,ar:ballester2015} have also reported results derived from P-band data in tropical forests which can be explained by anisotropic effects in wave propagation, whereas in \cite{ar:tello2018} a significant progress in this line has been made as horizontal and vertical forest structure indices were derived from SAR tomography at L-band. Likewise in \cite{ar:bondur2019} a new polarization signature was proposed for assessing the spatial variations in radar backscattering in boreal forests located in the coastal area of Lake Baikal within the Transbaikal conifer forests ecoregion. This approach provides evidences on the non-homogeneous structure of boreal forest cover sensed at L- and C-band through the assessment of the fluctuations of backscattered radar signals \cite{ar:bondur2008}.


%
%

More recently, the Random Volume over Ground (RVoG) model framework \cite{ar:treuhaft,ar:treuhaft_siqueira} has been extended in \cite{ar:zhu2023} by integrating the subaperture decomposition approach. This allows to increase the observation space on the basis of the different ground-to-volume ratios observed at different azimuthal angles and low frequencies. Consequently, authors in \cite{ar:zhu2023} devised a dual-baseline InSAR inversion approach aimed at generating digital terrain, surface, and canopy height models for both boreal and tropical forest at P-band.


Together, all the works reviewed above agree in emphasising that radar signatures from vegetation canopies may vary significantly during the SAR azimuthal coherent integration of backscattered pulses, thus violating the common assumption of the random volume morphology in radar-based vegetation studies. This is a potential source of error for parameter retrieval as their estimation is based on polarimetric and/or interferometric features. Therefore, this study further investigates the presence of these potential heterogeneities and/or anisotropic effects in volume backscattering \cite{ar:bondur2008,ar:garestier08_p,ar:dalessandro2013} by examining the azimuthal-dependent PolSAR signatures. To this aim one can resort to a number of polarimetric features that can be calculated at different azimuth look angles along the synthetic aperture. One possible option is to apply a polarimetric decomposition as in \cite{ar:ferro-famil2003} where, as mentioned above, the H-$\alpha$ decomposition was employed. However, that analysis was focused on agricultural fields where the entropy values were mostly 0.7 or below. Therefore, additional analyses in higher entropy scenarios (i.e. midterm or advanced phenological stages in crops or forests) must be dealt with an alternative strategy.

The purpose of this study is to further examine these scattering phenomena by means of the subaperture decomposition \cite{ar:ferro-famil2003} and the 3-D Barakat degree of polarisation \cite{ar:barakat}. Also, a discussion on some implications with respect the PolSAR techniques commonly applied is presented. To this aim, fully polarimetric radar scattering measurements from different scenarios have been analysed: 1) multi-frequency indoor acquisitions from two 1.8 m high vegetation samples; 2) P-band airborne data gathered from the boreal forest considered in the BioSAR2008 campaign; 3) ALOS-PALSAR-1 data from tropical forest in Paracou area, French Guiana.

The present work has been organised in the following way. Section \ref{s:met} describes the methodology and the theoretical background employed in the present analysis. The experimental datasets selected are described in Section \ref{s:data}. Section \ref{s:an} presents the results and a detailed analysis on them. A discussion on the implications and limitations of the present study is provided in Section \ref{s:disc} and the final conclusions will be drawn in Section \ref{s:conc}.

\section{Methodology}
\label{s:met}

In this study we have employed two different tools: 1) the 3-D Barakat degree of polarisation \cite{ar:barakat} and 2) the subaperture decomposition \cite{ar:ferro-famil2003}.

The 3-D Barakat degree of polarisation is defined as a function of the fully polarimetric coherency matrix $\mathbf{T}$ as follows \cite{ar:dey2021}:

\begin{equation}
m_{FP} = \sqrt{1 - \frac{27 |\mathbf{T}|}{tr^3(\mathbf{T})}}
\end{equation}
where $|\mathbf{T}|$ is the determinant and $tr(\mathbf{T})$ represents the trace of $\mathbf{T}$. A partially polarised wave is characterised by $m_{FP}$ between 0 and 1. The extreme behaviours correspond to a fully polarised wave represented by $m_{FP}=1$ whereas a totally unpolarised wave is described by $m_{FP}=0$. 

In \cite{ar:dey2021} it is proposed to retrieve the volume scattering mechanism intensity $P_v$ as a fraction of the total power (i.e. the $Span$) in the following way:

\begin{equation}
P_v = (1 - m_{FP})\cdot Span
\label{eq:pv}
\end{equation}

Therefore the factor $1-m_{FP}$ can be seen as an indicator of the contribution of the depolarising effects in a distributed target. When it comes to the monitoring of high entropy targets where a high randomness is expected, our hypothesis in this work is that this factor can be also used to detect anisotropic behaviour in certain pixels not fulfilling the random volume assumption. 

To give support to this idea we have computed and compared the entropy, $1-m_{FP}$ and the Radar Vegetation Index (RVI) \cite{pro:rvi} from two vegetation samples of small fir trees and maize measured under controlled conditions. Experiments were conducted at the European Microwave Signature Laboratory (EMSL) for a wide range of frequencies (both samples are 1.8 m high). See Section \ref{s:data} for a description of the data sets and references on them. 

\begin{figure*}[h!]
 \centering
\begin{tabular}{c}
  \includegraphics[width=14cm]{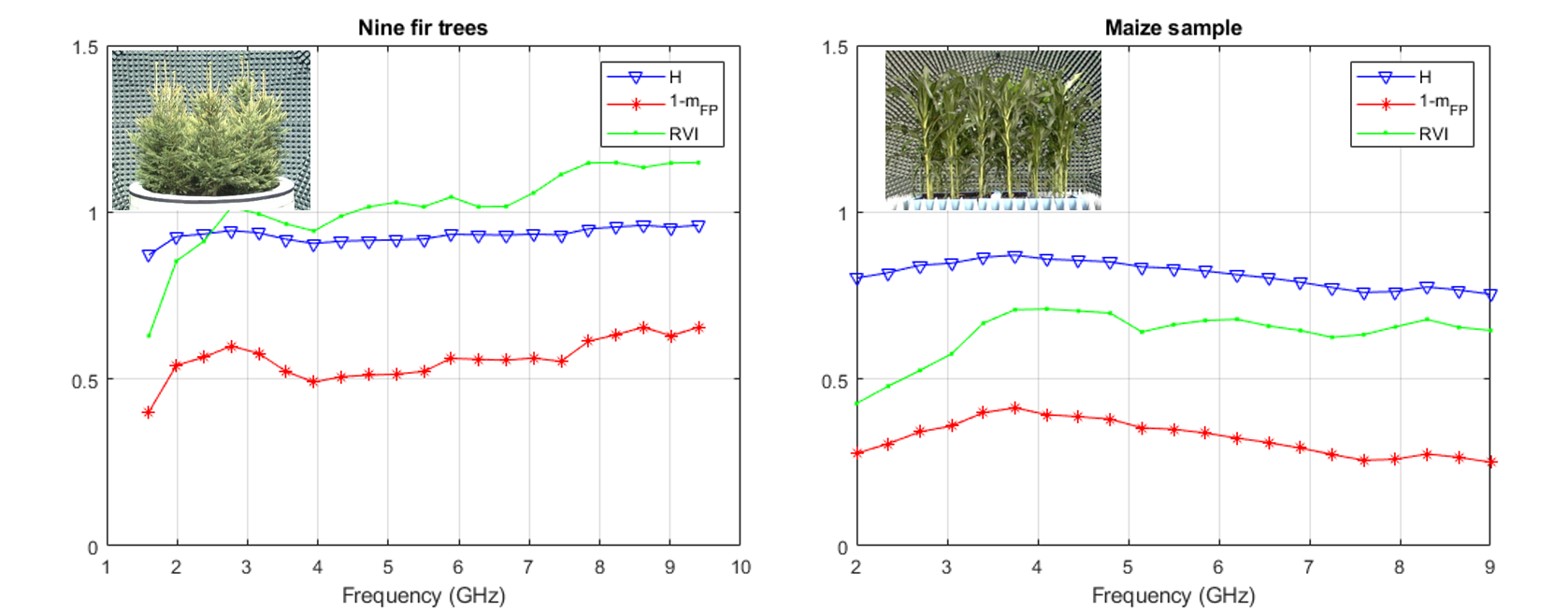}
\end{tabular}
\caption{\small Variation of H, $1-m_{FP}$ and RVI as a function of frequency for a fir trees cluster (left plot) and 6x6 maize plants (right plot). Experiments were conducted in the EMSL (see Section \ref{s:data} for details).}
\label{f:depol}
\end{figure*}

Figure \ref{f:depol} illustrates the quantitative and qualitative differences among H, $1-m_{FP}$ and RVI to characterise target randomness in radar signatures. It is noted that RVI should ideally vary between zero (bare surfaces) and one (dense vegetation cover). However, it can also reach values higher than one. Recently, two improved versions of RVI have been proposed in \cite{ar:szigarski} by adding information retrieved from the SMAP radiometer and, thus, increasing the robustness of polarimetric radar indices for vegetation mapping.

In case of fir trees (left plot in Figure \ref{f:depol}) the entropy exhibits stable and very high values ranging from 0.87 to 0.96. The RVI starts with 0.63 at low frequencies (i.e. L-band) but rapidly saturates to 1 (with small variations) and reaches values up to 1.15 at higher frequencies. On the other hand, the $1-m_{FP}$ factor shows higher dynamics ranging from 0.4 to 0.65. Overall, according to this data, the information that can be retrieved by both H and RVI indicators is limited to qualitative aspects, i.e. the target is characterised as a random volume or dense vegetation for the whole frequency range except at low frequencies. On the other hand, the $1-m_{FP}$ factor represents an approximate increasing trend of the volume scattering power as a function of frequency ranging from 40 to 65\% of the total backscattering power. This outcome is consistent with the increasing isotropic behaviour of the fir trees whose needles and short branches exhibit a dominant radar signature as the wavelength becomes shorter (i.e. a random volume).

For the maize sample (right plot in Figure \ref{f:depol}) all three indicators yield values associated to a lower level of randomness in comparison with the fir trees. Entropy varies between 0.76 and 0.87 revealing an upward and downward pattern and taking the lowest values at higher frequencies. RVI ranges from 0.42 to 0.71 from 2 to 3.5 GHz and then remains relatively stable around 0.7. The $1-m_{FP}$ factor seems again to provide the higher variability for the whole frequency interval, showing the same upward and downward pattern as for the entropy and varying from 0.25 to 0.41. As shown for this particular sample, the RVI is only sensitive up to 3.5 GHz whereas both the entropy and $1-m_{FP}$ exhibit a decreasing trend towards a less randomly polarised wave at medium and higher frequencies. It is pointed out that this behaviour is in agreement with the anisotropic scattering due to vertically oriented structures (i.e. the so-called oriented volume) which become dominant in the radar response. This outcome is supported by \cite{ar:cloude_letterpct} where the same dataset was employed and analysed by means of the coherence tomography technique \cite{ar:coh_tomog}. 

Overall, these observations suggest that the $1-m_{FP}$ factor outperforms H and RVI in terms of quantitative assessment of depolarising effects of vegetation canopies. Therefore, we will based our analysis on the assessment of the variation of the $1-m_{FP}$ factor as the azimuth look angle changes.

\subsection{Subaperture Decomposition}

\begin{figure*}[h!]
 \centering
\begin{tabular}{c}
  \includegraphics[width=8cm]{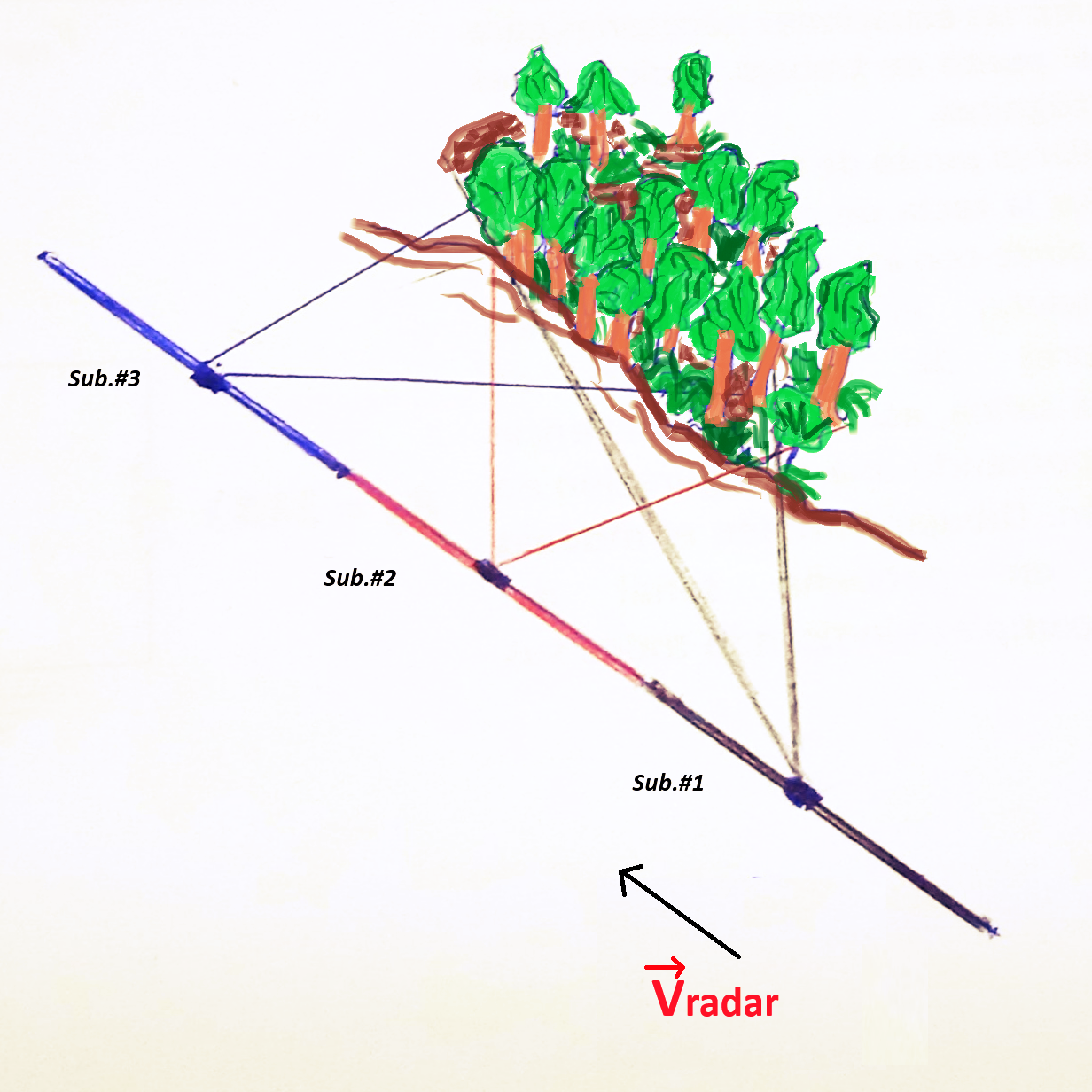}
\end{tabular}
\caption{\small Diagram illustrating the subaperture decomposition concept.}
\label{f:scheme}
\end{figure*}

The SAR imaging process consists of a high amount of low-resolution echoes from a scene collected by the moving sensor at azimuthal angles which are combined to create a full resolution radar image. As a result, each pixel in a SAR image represents an observation of an area across a specific range of angles constrained by the azimuth antenna pattern. The Doppler spectrum arising as a consequence of this time-varying azimuthal observation of the scene can be exploited to analyse the potential backscattering variations along the azimuth angle \cite{ar:ferro-famil2003}. Figure \ref{f:scheme} provides a simplified scheme of the subaperture decomposition concept by using three subapertures (indicated in black, red and blue colors).

Equation (\ref{eq:dopp}) provides the link among the Doppler frequency $f_{D}$, the sensor velocity $V_{radar}$, the azimuth look angle $\phi$ and the radar wavelength $\lambda$ \cite{b:curlander}:

\begin{equation}
f_{D} = \frac{2 V_{radar}}{\lambda}\, \sin\phi \label{eq:dopp}
\end{equation}

The relationship between azimuth resolution $\delta_{az}$ and the azimuth observation interval $\Delta\phi$ is as follows \cite{b:curlander}: 

\begin{equation}
\Delta\phi = \frac{\lambda}{2\delta_{az}} \label{eq:deltaphi}
\end{equation}

This approach is performed by first applying the Fourier transform along azimuth dimension. Then, in the Doppler spectral domain a SAR signal for a particular range bin $r$ is expressed as follows \cite{b:lee}: 

\begin{equation}
S(f_D,r) = S_{t}(f_D,r)H(f_D,r)W(f_D,r) 
\end{equation}
where $S_{t}(\cdot)$, $H(\cdot)$ and $W(\cdot)$ represent the Fourier transforms of the
transmitted signal, the scene coherent reflectivity and the focusing reference function (both included in $H(\cdot)$) and the weighting function, respectively. An average sidelobe reduction window can be recovered to correct for the spectral imbalance $W(f_D,r)$ in the full resolution image \cite{ar:ferro-famil2003}. Once the weighting function has been compensated, the third step is the splitting of the original Doppler spectrum into several sublooks (with or without overlapping) represented by a reduced Doppler bandwidth around a particular Doppler frequency $f_{D0}$ by applying the windowing function $G(f_D-f_{D0},r)$:

\begin{equation}
S(f_D-f_{D0},r) = S_{t}(f_D,r)H(f_D,r)G(f_D-f_{D0},r) 
\end{equation}

It is noted that function $G(f_D-f_{D0},r)$ accounts also for the sidelobe reduction in each subaperture spectra before they are converted back to spatial domain by an inverse Fourier transform. This procedure leads to a number of lower resolution SAR images according to the selected amount of subapertures employed. It must noted, however, that results differ depending on the number of subapertures and the degree of overlapping among them in the spectral domain \cite{th:sanjuan,ar:sanjuan2015} and this in turn is directly related to the resolution loss \cite{ar:garestier08_p}.


\section{Test sites and Datasets}
\label{s:data}

The three datasets employed in this study are briefly described next.

\subsection{Multi-frequency indoor data}

\begin{figure*}[h!]
 \centering
\begin{tabular}{cc}
  \includegraphics[width=6.85cm]{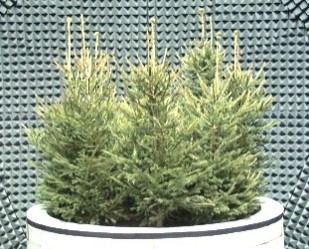}  &   \includegraphics[width=8cm]{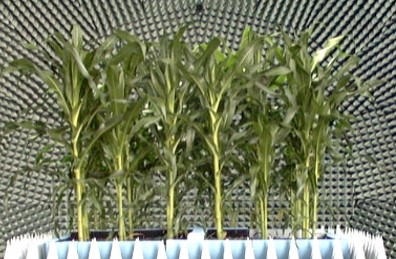}\\
	a) & b)
\end{tabular}
\caption{\small Photographs of the vegetation samples measured at the EMSL. a) Nine fir trees (1.8 m high) planted in a 2.4 m diameter round container; b) 6 x 6 young maize plants (1.8 m high) uniformly planted in a square container of side length 2 m.}
\label{f:vgm}
\end{figure*}

First, we have employed indoor polarimetric data from test vegetation samples (i.e. small fir trees and maize-see Figure \ref{f:vgm}) acquired under controlled conditions at the European Microwave Signature Laboratory (EMSL) \cite{web:emsl}. These datasets have been widely employed for vegetation radar scattering studies \cite{b:mitesis,th:lluis,ar:ovog,ar:cloude_letterpct}. The antenna beamwidth uniformly illuminates the vegetation sample and the system is operated in a stepped-frequency mode in 10 MHz steps ranging from 1.6 to 9.4 GHz for the fir trees sample and from 2 to 9 GHz for the maize.  Data acquisitions cover the whole azimuth interval at 5$^\circ$ steps making available 72 measurement through the azimuth look angle. This set-up allows an analysis of the polarimetric signatures as a function of the azimuthal observation which is relevant for the purpose of the present study as the whole azimuthal span can be splitted into four quadrants (hereinafter also referred to as \emph{subapertures}) thus providing a way to analyse vegetation scattering variations depending on the azimuth look angle.

\subsection{P-band BioSAR 2008 - Boreal forest}

The P-band polarimetric SAR datasets acquired by DLR's E-SAR system during BioSAR 2008 campaign have been also employed \cite{b:biosar2008}. This campaign was carried out in the Krycklan river catchment in northern Sweden, an area characterised by hilly terrain with elevation ranging from 100 to 400 m and surface slopes of up to 14$^\circ$ within the monitored forest stands.

\begin{figure*}[h!]
 \centering
\begin{tabular}{c}
  \includegraphics[width=11cm]{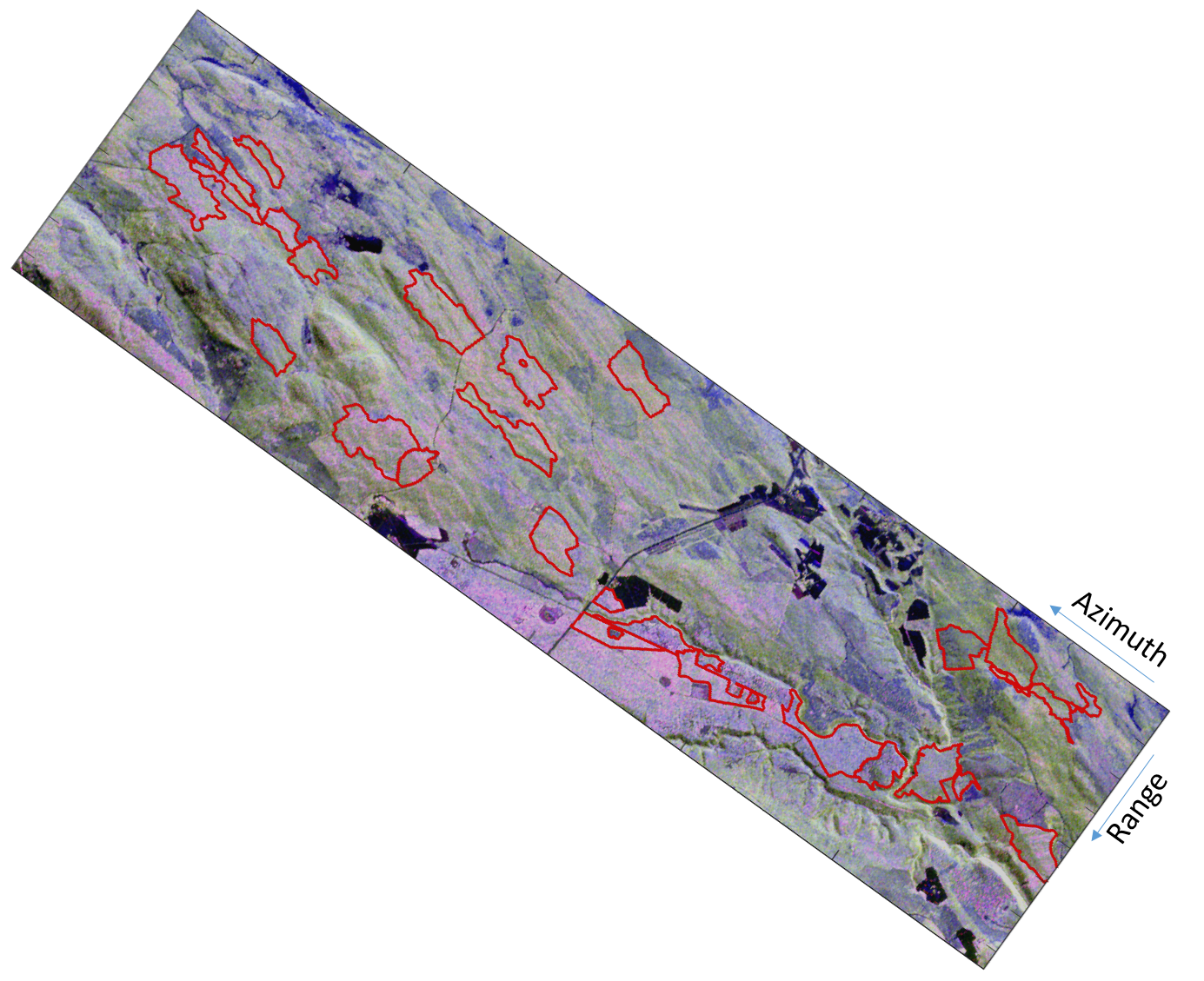}
\end{tabular}
\caption{\small BioSAR2008 test site. Monitored stands are indicated in red color. See \cite{b:biosar2008} for further details.}
\label{f:vgm18}
\end{figure*}

All P-band polarimetric features have been computed by employing a 9$\times$9 boxcar filter.

\subsection{ALOS PALSAR-1 - Tropical forest}

\begin{figure*}[h!]
 \centering
\begin{tabular}{c}
  \includegraphics[width=11cm]{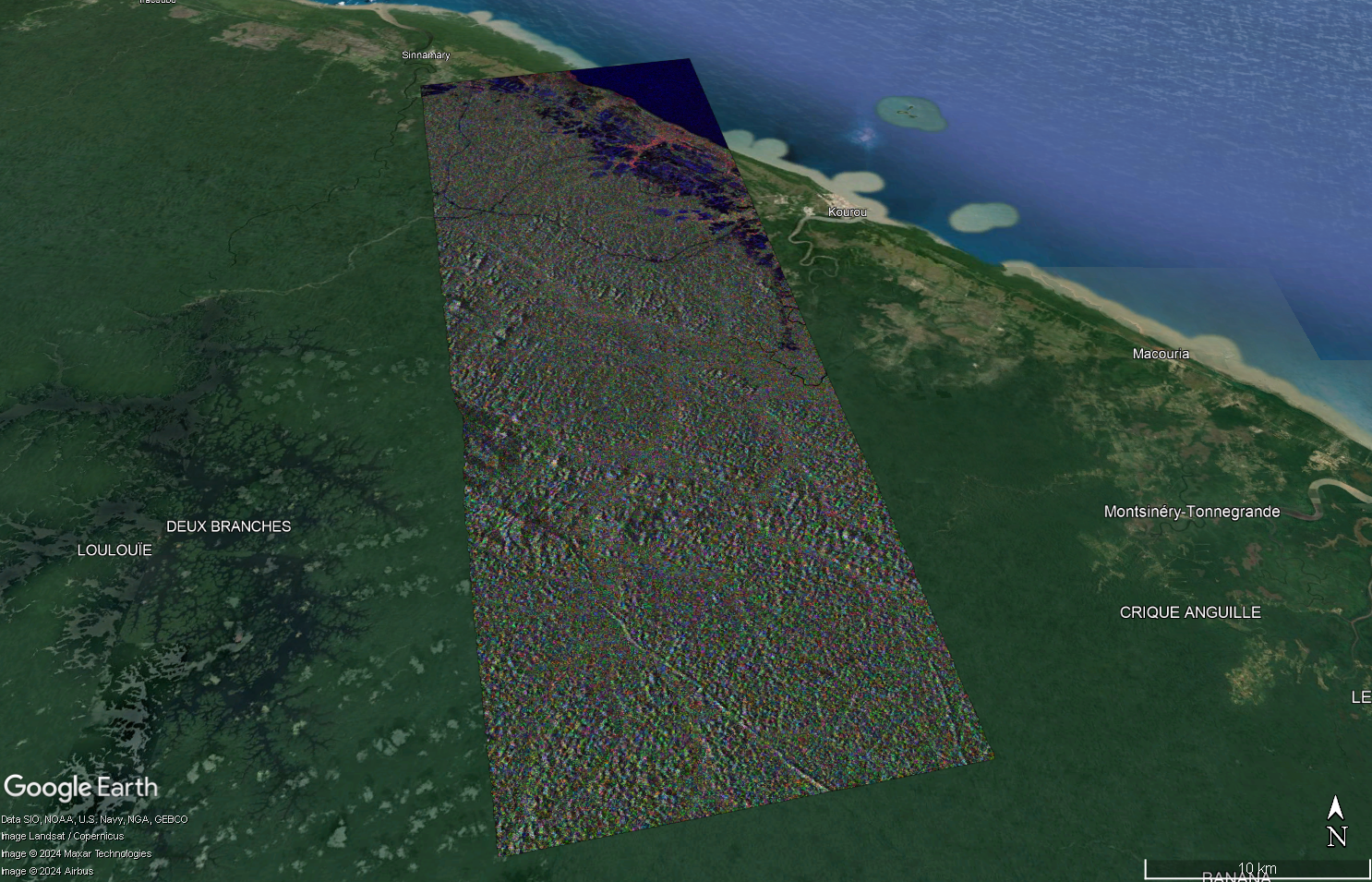}
\end{tabular}
\caption{\small ALOS PALSAR-1 SAR composition from the Paracou test site (product name ALPSRP176360090-L1.1). Dataset downloaded from the Alaska SAR Facility website (\texttt{https://search.asf.alaska.edu/})}
\label{f:paracou}
\end{figure*}

The selected product is an ALOS PALSAR Level 1.1 Single Look Complex (SLC) image. It was downloaded from the Alaska SAR Facility website (\texttt{https://search.asf.alaska.edu/}). The product ( ALPSRP176360090-L1.1) covers the tropical forest in the Paracou area, French Guiana, a well-known test site as it was monitored in the framework of the TropiSAR campaign conducted in 2009 \cite{b:tropisar09}. Terrain exhibits gentle topographic variations of about 30 m. Figure \ref{f:paracou} shows the Paracou test site.

In order to avoid the possible effect of slopes terrain, we have employed the Google Earth tool aimed at providing the elevation profile and the terrain slopes to focus our analysis in areas where the slopes are within the range $\pm$2\% which corresponds to a $\pm$1.15$^\circ$ interval. All L-band polarimetric features have been computed by employing a 9$\times$9 boxcar filter.

\section{Analysis}
\label{s:an}

\subsection{Multi-frequency indoor data}
\label{s:indoor}

\begin{figure*}[h!]
 \centering
\begin{tabular}{c}
  \includegraphics[width=10cm]{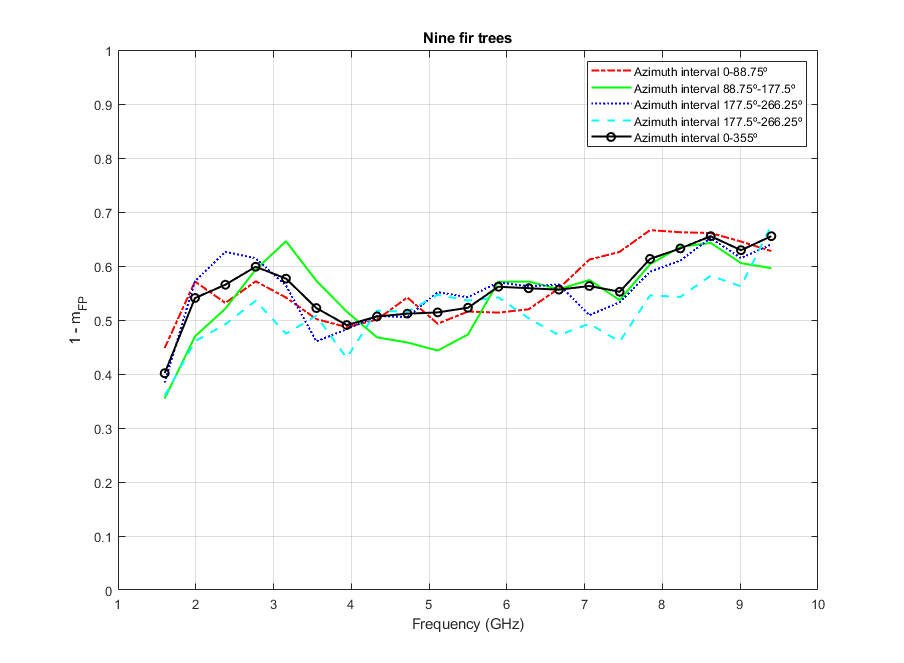}
\end{tabular}
\caption{\small Variation of $1-m_{FP}$ as a function of frequency for the fir trees cluster for reduced azimuth intervals of approximately 90$^\circ$ (different line styles without markers) and for the whole azimuth span (line with circles).}
\label{f:vgm12_m}
\end{figure*}
\begin{figure*}[h!]
 \centering
\begin{tabular}{c}
  \includegraphics[width=10cm]{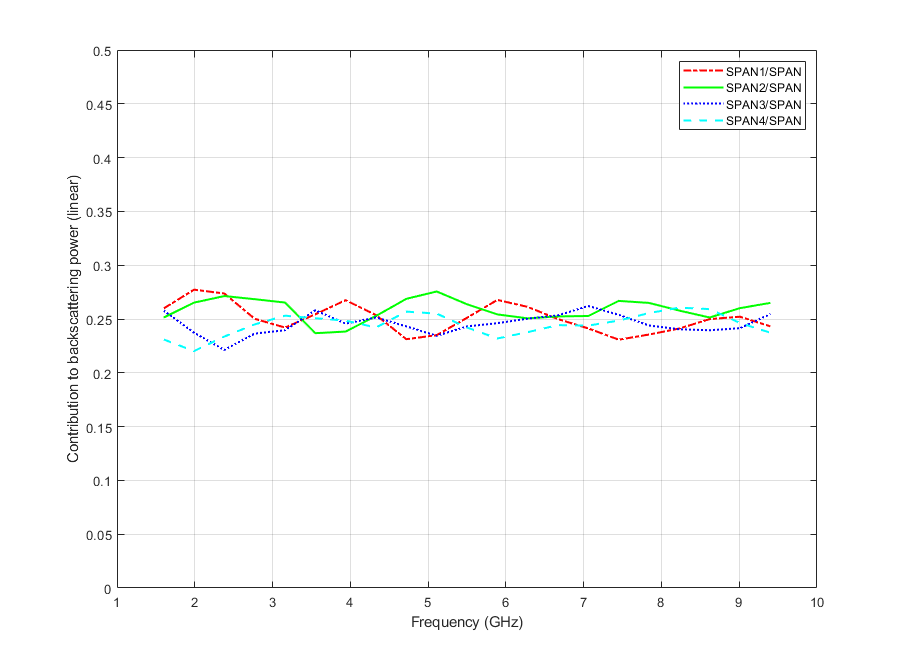}
	\end{tabular}
\caption{\small Variation of the contribution of the total power for each azimuth interval, $Span_i$, relative to the total power accounting for the whole azimuth span as a function of frequency for the fir trees.}
\label{f:vgm12_s}
\end{figure*}

Figure \ref{f:vgm12_m} shows the variation of $1-m_{FP}$ as a function of frequency for the four quadrants of the whole azimuthal span and the corresponding $1-m_{FP}$ values obtained by averaging all samples acquired by covering the whole azimuth interval (shown before in Figure \ref{f:depol} and repeated here for convenience). Most of $1-m_{FP}$ values, either for partial or the whole azimuth span, exhibit levels very close to or higher than 0.5 (except at L-band and around 4 and 5 GHz for two particular subapertures). In terms of backscattered power this means that the depolarising effects dominate the radar response. However, the $1-m_{FP}$ plots for all four azimuth quadrants clearly show different behaviours for different frequency bands which translates into different contributions of the diffuse backscattering power to the total power of the particular subaperture. At L-band there is a 9\% difference between the first and both the second and fourth subapertures. The highest differences have been found to be 17\% at about 3 GHz between the second and fourth subapertures and also at 7.5 GHz between the first and the fourth azimuth intervals. It must be noted here that speckle noise has been properly accounted for as the number of samples averaged is 162 (i.e. 9 samples within a bandwidth of 1 GHz around the central frequency and 18 azimuthal samples per frequency) and, hence, the observed variations must be attributed to different scattering behaviours.

Together with the depolarising effects substantiated in $1-m_{FP}$ the contributions of the total backscattered power of each individual subaperture to the total span of the whole image have been also computed. It is the ratio $Span_i/Span$, being $Span_i$ the span of the subaperture $i$ and $Span$ the total backscattered power for the whole azimuth illumination. They are shown in Figure \ref{f:vgm12_s}. As the full azimuth interval has been equally split into four quadrants, all four ratios would be ideally 0.25 whenever the assumption of an isotropic distributed target is observed. As shown, the highest differences appear at S-band between 1st-2nd and 3rd-4th quadrants but the ratio values are within the 22-28\% interval which is pretty close to the expected 25\% ratio.

\begin{figure*}[h!]
 \centering
\begin{tabular}{c}
  \includegraphics[width=9cm]{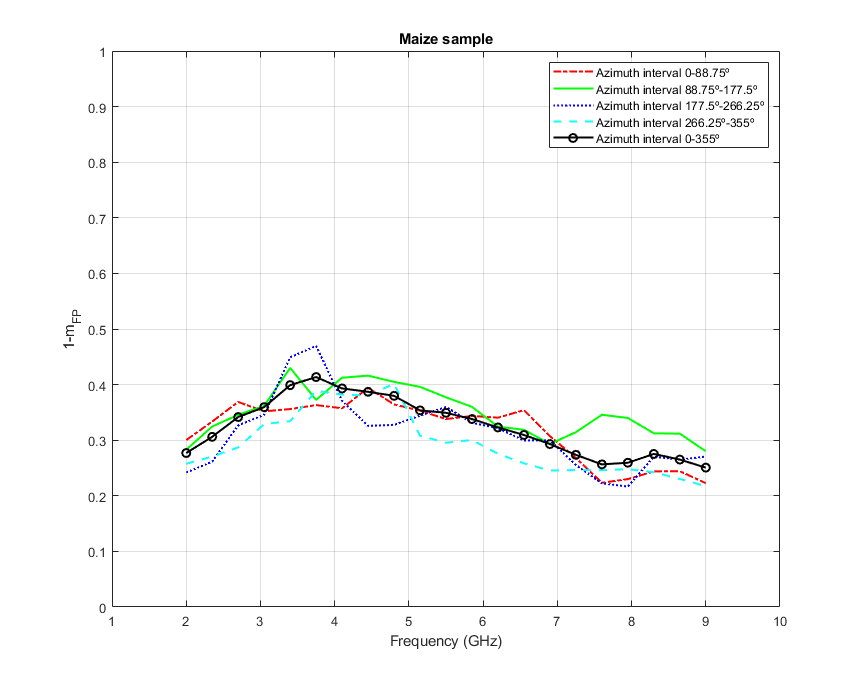}
\end{tabular}
\caption{\small Variation of $1-m_{FP}$ as a function of frequency for the maize sample for reduced azimuth intervals of approximately 90$^\circ$ (different line styles without markers) and for the whole azimuth span (line with circles).}
\label{f:vgm18_m}
\end{figure*}

\begin{figure*}[h!]
 \centering
\begin{tabular}{c}
  \includegraphics[width=9cm]{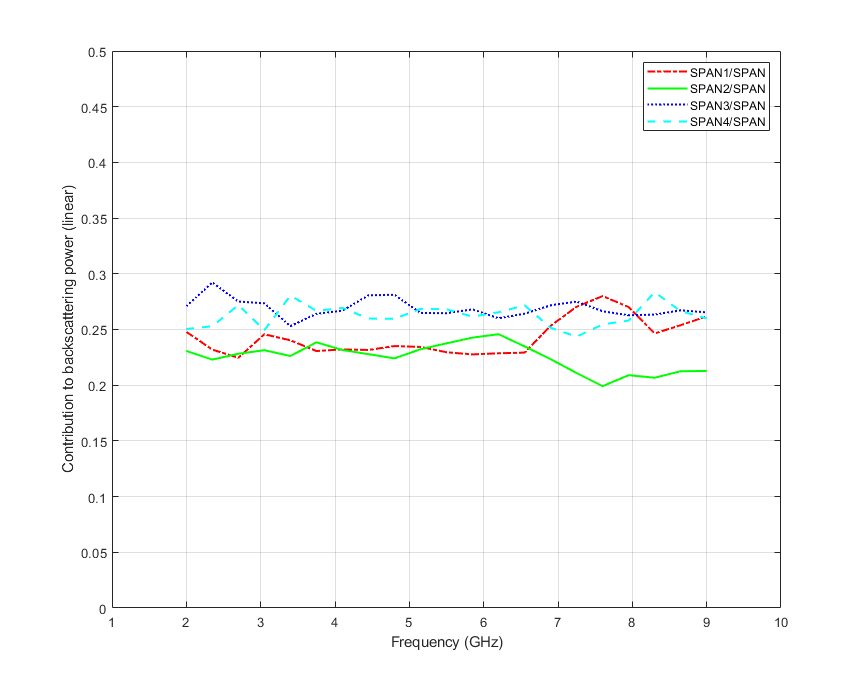}
	\end{tabular}
\caption{\small Variation of the contribution of the total power for each azimuth interval, $Span_i$, relative to the total power accounting for the whole azimuth span $Span$ as a function of frequency for the maize sample.}
\label{f:vgm18_s}
\end{figure*}

The same analysis has been carried out for the maize sample and the corresponding plots are shown in Figures \ref{f:vgm18_m} and \ref{f:vgm18_s}. This case shows a qualitative difference with respect the fir trees cluster. The $1-m_{FP}$ values are consistently below 0.5, being most of them below 0.4. Indeed, from 4 GHz onwards values follow a decreasing trend up to about 0.25 at higher frequencies. This behaviour indicates that polarised components become dominant in the radar signature which means that anisotropic effects occur in the wave propagation, despite the power ratios are also equally distributed among all subapertures (except for the second subaperture at higher frequencies which decreases to about 20\%). This observation is in agreement with the conclusions drawn in \cite{ar:cloude_letterpct} by employing the coherence tomography approach. It is noted that the differences among $1-m_{FP}$ for all four subapertures reach values about 10\% (up to 12.5\% at 8 GHz). These values do not reach the ones for the fir trees but certainly they represent noticeable differences regarding the depolarising effects as a function of the azimuth look angle.

\subsection{P-band BioSAR 2008 - Boreal forest}
\label{s:biosar}

\begin{figure*}[h!]
 \centering
\begin{tabular}{c}
  \includegraphics[width=11cm]{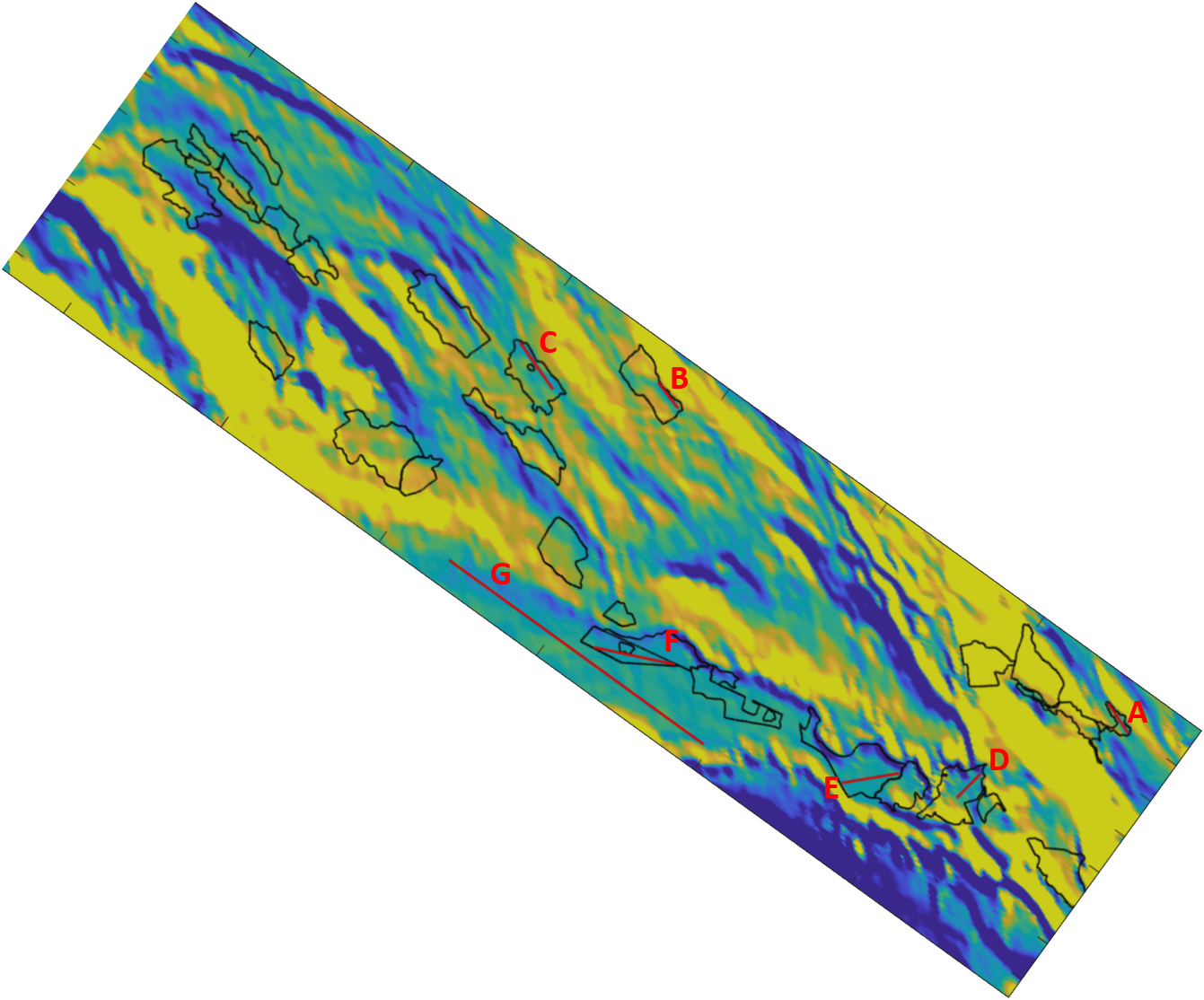}
\end{tabular}
\caption{\small Slope map of the BioSAR2008 test site. Monitored stands during the BioSAR2008 campaign are drawn in black and the transects analysed are indicated in red color and capital letters.}
\label{f:biosar_slope}
\end{figure*}

The slope map provided in the BioSAR2008 campaign is displayed in Figure \ref{f:biosar_slope} where the monitored stands are overlaped. The pixels with slopes outside the interval $\pm$5$^\circ$ have been saturated and are shown in yellow and dark blue colors on the map. For our analysis we have focused on zones where terrain slopes effects are negligible so we have selected transects whose slopes range between -2.5$^\circ$ and 1.25$^\circ$. These transects are indicated on the slope map with red color lines and letters.

\begin{figure*}[h!]
 \centering
\begin{tabular}{cc}
  \includegraphics[width=8cm]{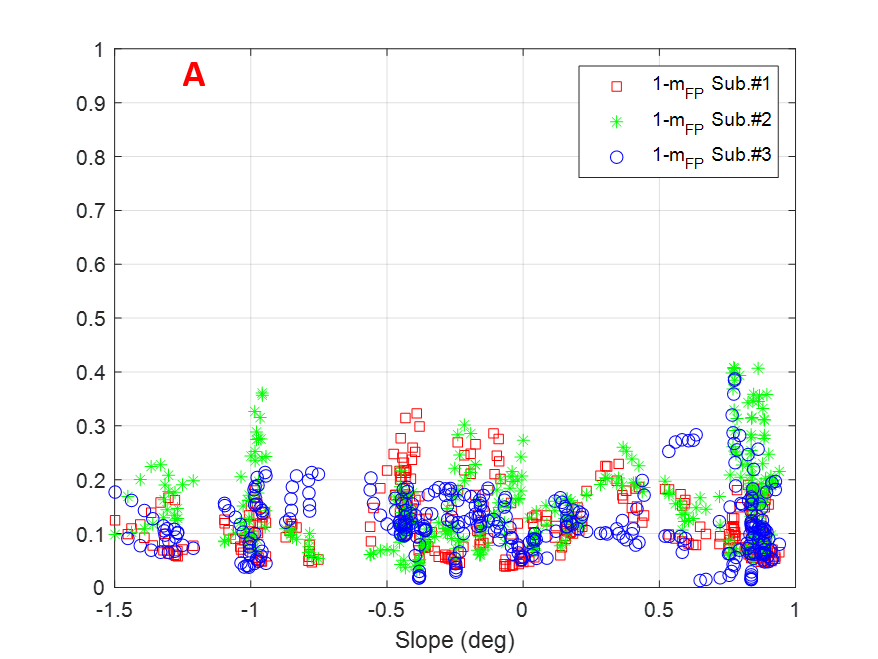}  &   \includegraphics[width=8cm]{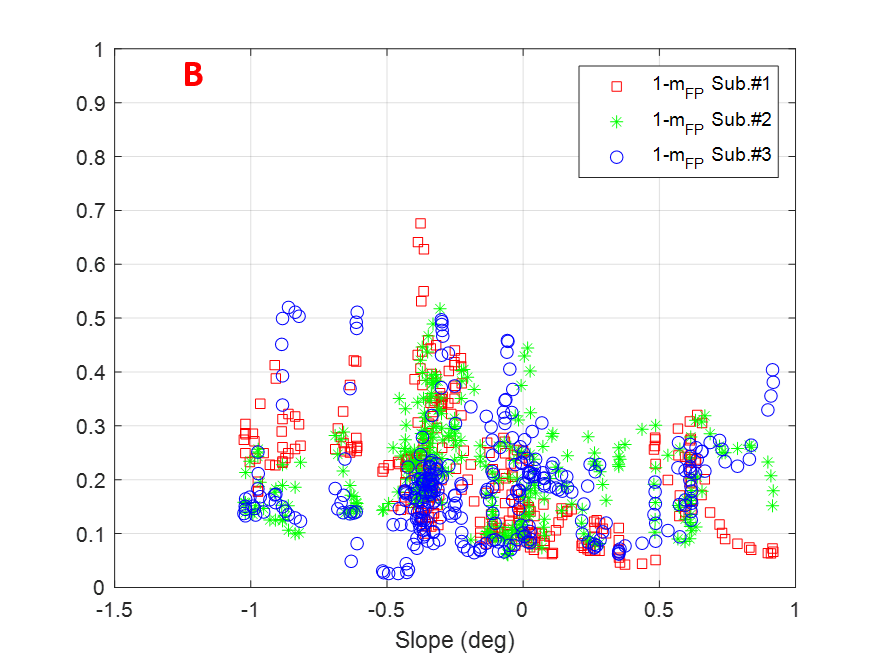}\\
	\includegraphics[width=8cm]{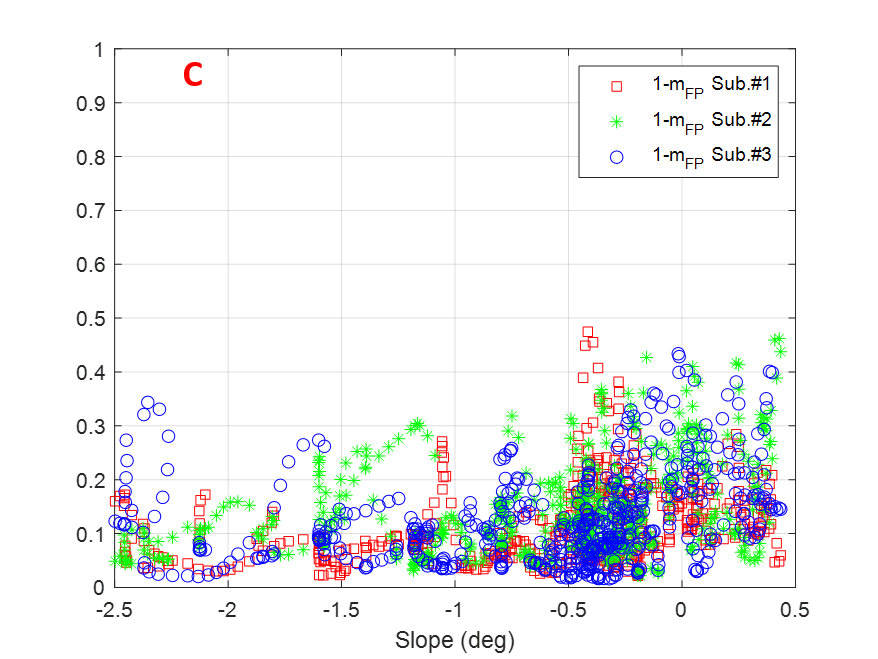}  &   \includegraphics[width=8cm]{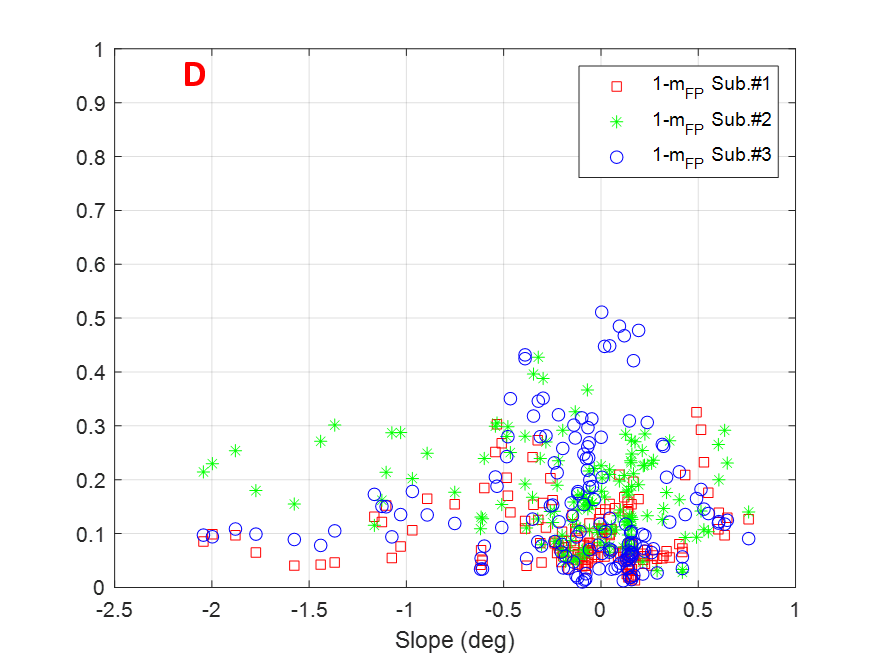}\\
	\includegraphics[width=8cm]{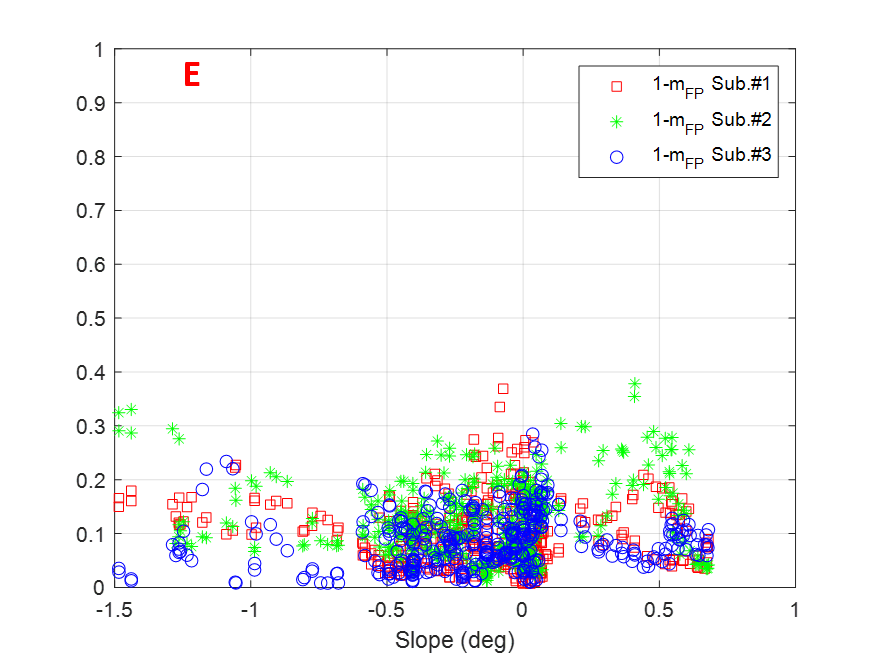}  &   \includegraphics[width=8cm]{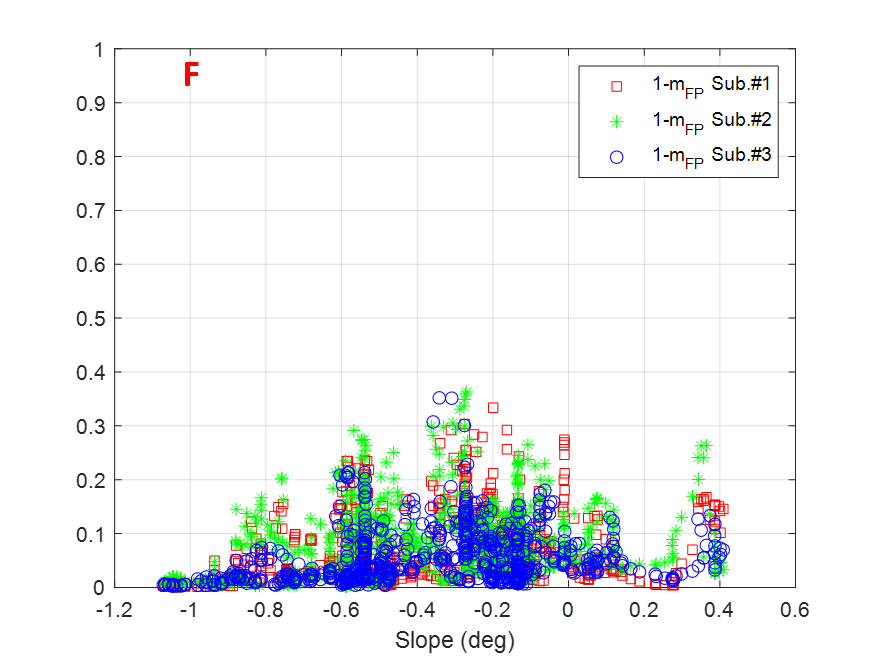}
\end{tabular}
\caption{\small Values of $1-m_{FP}$ for all three subapertures as a function of terrain slopes for the transects selected and indicated with letters A-F in Figure \ref{f:biosar_slope}.}
\label{f:transects}
\end{figure*}

Figure \ref{f:transects} displays the $1-m_{FP}$ factor for individual subapertures as a function of terrain slopes for the transects selected and indicated with letters A-F in Figure \ref{f:biosar_slope}. In general, the observed values are mostly limited within the $[0,0.3]$ range, therefore the diffuse scattering contribution for individual subapertures can be at most 30\% of the total backscattered power for a particular azimuth look angle. This is not a dominant volume scattering which agrees well with SAR tomography studies \cite{ar:frey} reporting a dominant polarised ground signature at P-band. Also it is observed that there is no clear pattern when comparing the $1-m_{FP}$ values among subapertures, thus suggesting that anisotropic effects arise depending upon the azimuth look angle and that they can be present in more than one subaperture (see transect A for slopes around -1$^\circ$ or 0) or transect B for -0.5$^\circ$ slopes. Transects A, B, C and D exhibit also some parts where depolarising effects in some subapertures are more noticeable reaching values higher than 0.3.

In addition to the analysis of the $1-m_{FP}$, we have also applied the stationarity test proposed in \cite{ar:ferro-famil2003} where a Maximum Likelihood (ML) ratio test is designed to discriminate between anisotropic and isotropic scattering signatures as follows:

\begin{equation}
\Lambda = \left(\frac{\prod_{i=1}^{Nsub}\mathbf{T_i}}{\left[\left(\frac{1}{Nsub}\right)^{Nsub} \left|\sum_{i=1}^{Nsub} \mathbf{T_i}\right|\right]^{Nsub}}\right)^{N}\label{eq:lambda}
\end{equation}
In Eq.(\ref{eq:lambda}) $Nsub$ represent the number of subapertures, $\mathbf{T_i}$ is the coherency matrix for subaperture $i$, and $N$ is the number of samples for averaging. It is noted that this is an expression different to that provided in \cite{ar:ferro-famil2003} but totally equivalent.

According to \cite {ar:ferro-famil2003}, the $\Lambda$ parameter is expected to increase in natural covers due to the stationary spectral behaviour from distributed targets. Contrarily, $\Lambda$ will decrease for anisotropic pixels as the stationarity assumption is not fulfilled.

\begin{figure*}[h!]
 \centering
\begin{tabular}{c}
  \includegraphics[width=16cm]{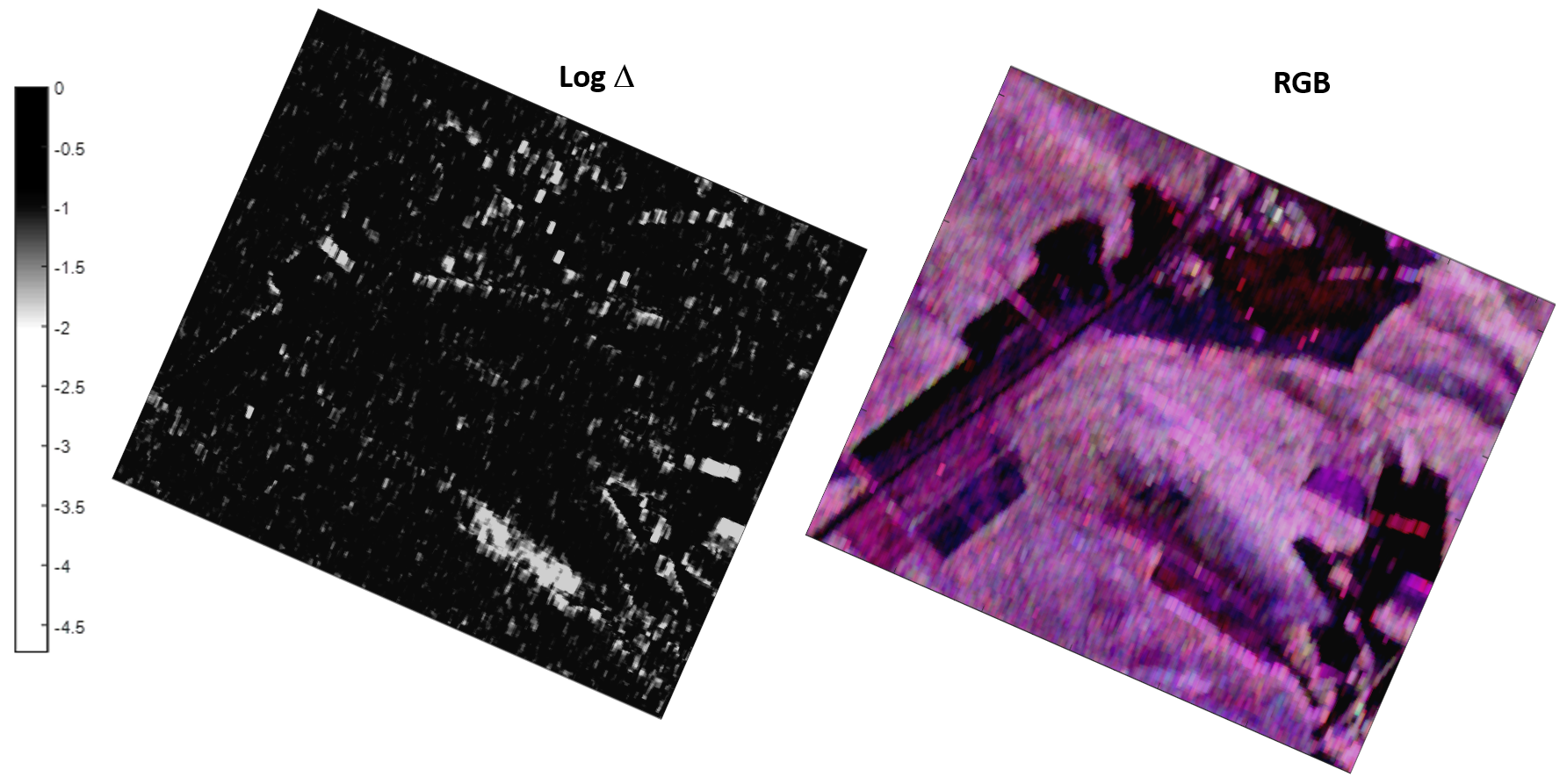}
\end{tabular}
\caption{\small Map of the ML ratio ($\log\Lambda$) for the BioSAR2008 test site corresponding to the area showed in the RGB map. In the $\log\Lambda$ map, black level represents isotropic pixels (mostly corresponding to natural covers) and white level indicates the presence of anisotropic structures.}
\label{f:lambda}
\end{figure*}

In order to better understand the correspondence between the $\Lambda$ parameter and the anisotropic-isotropic behaviour, we have computed the ML ratio for the zone located at near range where several anisotropic and man-made structures can be seen which are surrounded by natural areas (see the area close to the coast in Figure \ref{f:paracou}). Figure \ref{f:lambda} shows the results of $\Lambda$ in logarithmic scale (i.e. $\log\Lambda$) together with and RGB image in the lexicographic basis (R=HH, G=HV, B=VV). 

The $\log\Lambda$ is displayed in a grey scale image where black level corresponds to values from 0 to -1, gray levels represent a transition interval between -1 and -2, and white level ranges from -2 to the lowest $\log\Lambda$ value found in the image. All structures showing a preferred orientation are clearly visible in white and light grey shades whereas the isotropic behaviour is displayed in black. 

\begin{figure*}[h!]
 \centering
\begin{tabular}{c}
  \includegraphics[width=16cm]{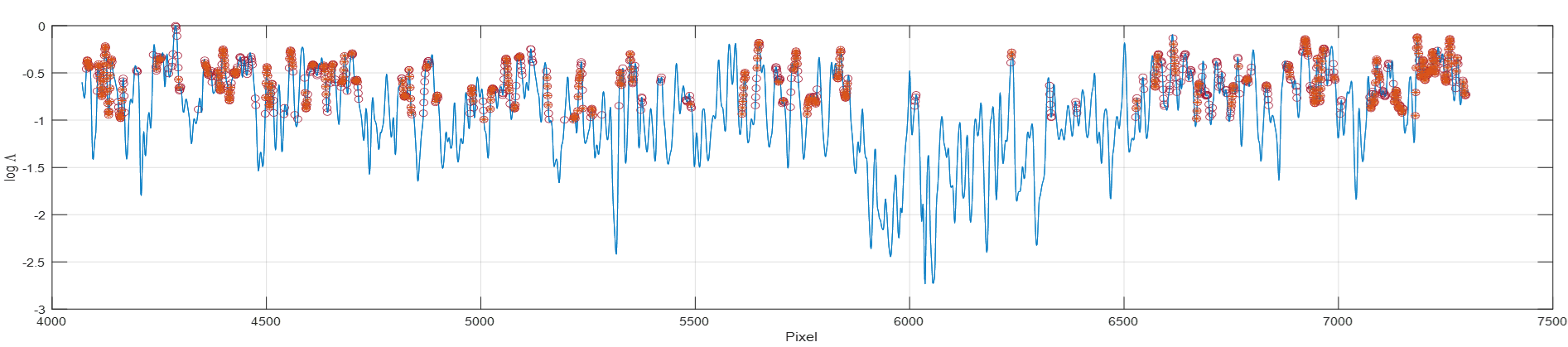}
\end{tabular}
\caption{\small Map of the ML ratio ($\log\Lambda$) for the transect G indicated in Figure \ref{f:biosar_slope}. Circles indicate pixels where a difference in $1-m_{FP}$ between any two subapertures is higher than 0.1 and where $\log\Lambda >$-1. Stars indicate pixels where a difference in $1-m_{FP}$ between any two subapertures is higher than 0.15 and where $\log\Lambda >$-1.}
\label{f:lambda_tran}
\end{figure*}

According to the previous results for this particular dataset, we set the threshold for discriminating between anisotropic and isotropic pixels in $\log \Lambda$=-1. By doing so, it seems possible to detect heterogeneities in forest covers. We have computed the ML ratio for a long transect (named G in Figure \ref{f:biosar_slope}) where slopes are within the $[-1.9^\circ, 1.25^\circ]$ range. Figure \ref{f:lambda_tran} shows this result in solid line. Circles indicate pixels where a difference in $1-m_{FP}$ between any two subapertures is higher than 0.1 and where $\log\Lambda >$-1. Stars indicate pixels where a difference in $1-m_{FP}$ between any two subapertures is higher than 0.15 and where $\log\Lambda >$-1. The former case represents 28.39\% of pixels for that transect whereas the latter case accounts for the 14.84\% of pixels. Therefore, despite the stationarity test yields levels associated to an isotropic medium, the differences among depolarising factors suggest that vegetation heterogeneities could be affecting the scattering signature in a significant amount of pixels.

\begin{figure*}[h!]
 \centering
\begin{tabular}{cc}
  \includegraphics[width=8cm]{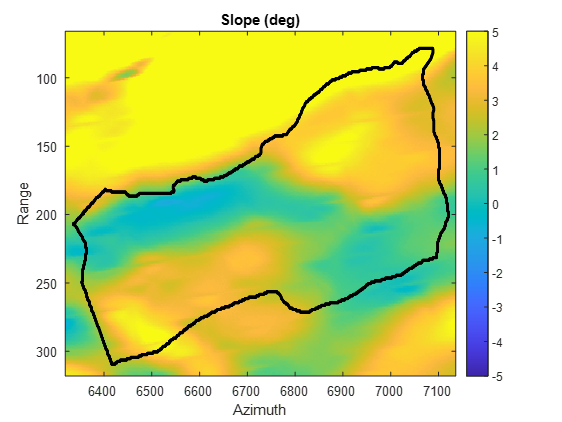}  &   \includegraphics[width=8cm]{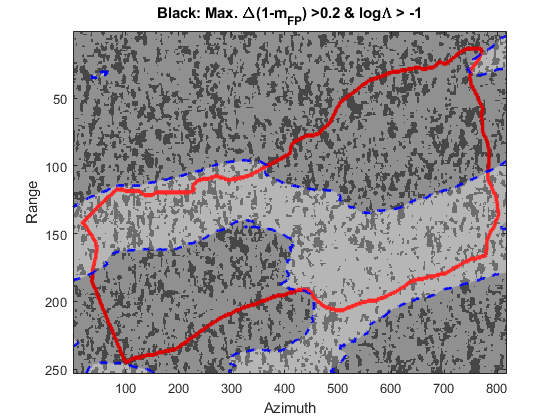}\\
	a) & b)
\end{tabular}
\caption{\small Area within BioSAR2008 test site corresponding to stand containing transect B indicated in Figure \ref{f:biosar_slope}. a) Map of terrain slope (saturated to $\pm5^\circ$) of the selected area. Black line represents the limits of the forest stand; b) Map of pixels (shown as darker patches) where $1-m_{FP}$ differences between any two subapertures is higher than 0.2 and where the stationarity hypothesis is fulfilled. Area inside the dotted blue line region are pixels where the slope terrain is within the $\pm$2$^\circ$ interval.}
\label{f:stand}
\end{figure*}

To support the previous interpretation based on the $1-m_{FP}$ differences and the $\log\Lambda$ indicator, we have  computed a $\log\Lambda$ map corresponding to the highest forest stand (the one which contains transect B in Figure \ref{f:biosar_slope} whose average height is 21.408 m \cite{b:biosar2008}). Figure \ref{f:stand}.a) displays the terrain slope of the selected area where the black line represents the limits of the forest stand. Figure \ref{f:stand}.b) shows a grey scale map where scattered darker patches show the pixels that simultaneously exhibit a $1-m_{FP}$ difference between any two subapertures higher than 0.2 and where the stationarity hypothesis is fulfilled (in our analysis, $\log\Lambda >$-1). Light grey shades inside the dotted blue line region are pixels where the slope terrain is within the $\pm$2$^\circ$ interval. We focus on this particular zone to avoid that polarimetric signatures get modified by slope terrain. The number of pixels fulfilling both conditions represents 25.05\% which means that, for this P-band dataset, one fourth of those pixels exhibit polarimetric signatures not compatible with the expected isotropic signature from forest covers. It is noted that if a 0.1 difference among subapertures $1-m_{FP}$ was considered, the number of pixels where potential anisotropic effects would appear increases to 58.71\%.

\begin{figure*}[h!]
 \centering
\begin{tabular}{c}
  \includegraphics[width=15cm]{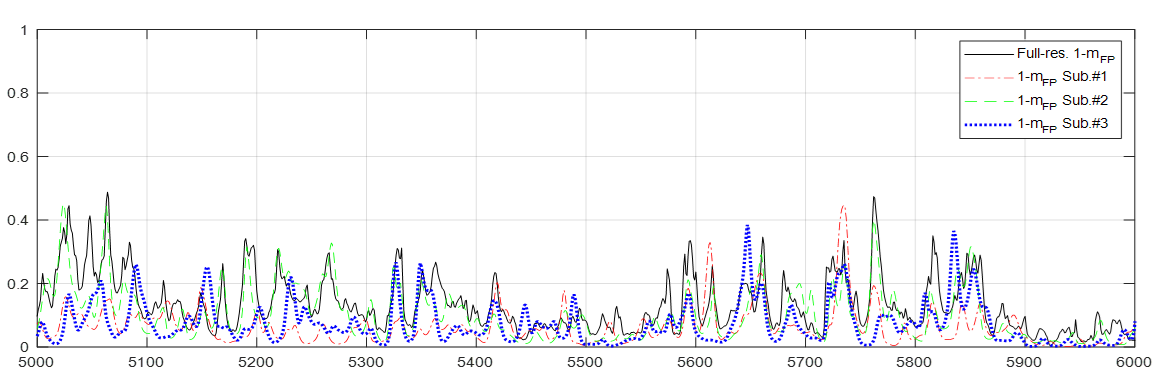} \\
	\includegraphics[width=15cm]{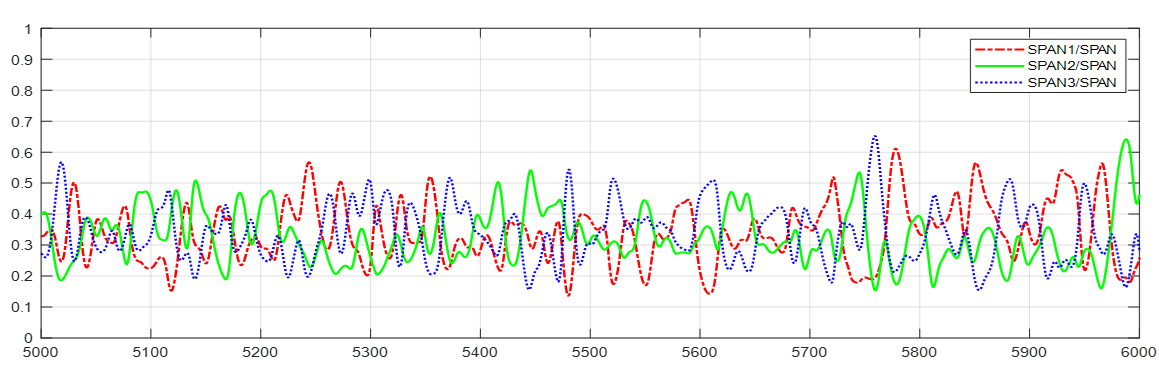}
\end{tabular}
\caption{\small Variation for each subaperture from pixel 5000 to pixel 6000 of transect G of (top image) $1-m_{FP}$ factor and (bottom image) the relative contribution of total backscattering with respect the span of the full resolution image.}
\label{f:tran}
\end{figure*}

For this dataset, this means that the relative contribution of the diffuse scattering to the total backscattering of an individual subaperture can noticeably vary. This leads to different balances between polarised and unpolarised components depending on the considered subaperture even though the ML ratio test characterise those pixels as isotropic targets. Therefore, this fact suggests that the assessment of anisotropic pixels in forests can be carried out on the basis of the information provided by $1-m_{FP}$. Additionally, the total backscattering power gathered by an individual subaperture, i.e. $Span_i/Span$, can also provide additional information, as seen next.

%

A zoomed area from pixel 5000 to pixel 6000 in transect G is displayed in Figure \ref{f:tran} where, for each subaperture, the $1-m_{FP}$ factor and the relative contribution of total backscattering are displayed. For comparison purposes, the full resolution $1-m_{FP}$ has been also plotted (black solid line) on the top image in Figure \ref{f:tran}.

\begin{figure*}[h!]
 \centering
\begin{tabular}{c}
  \includegraphics[width=16cm]{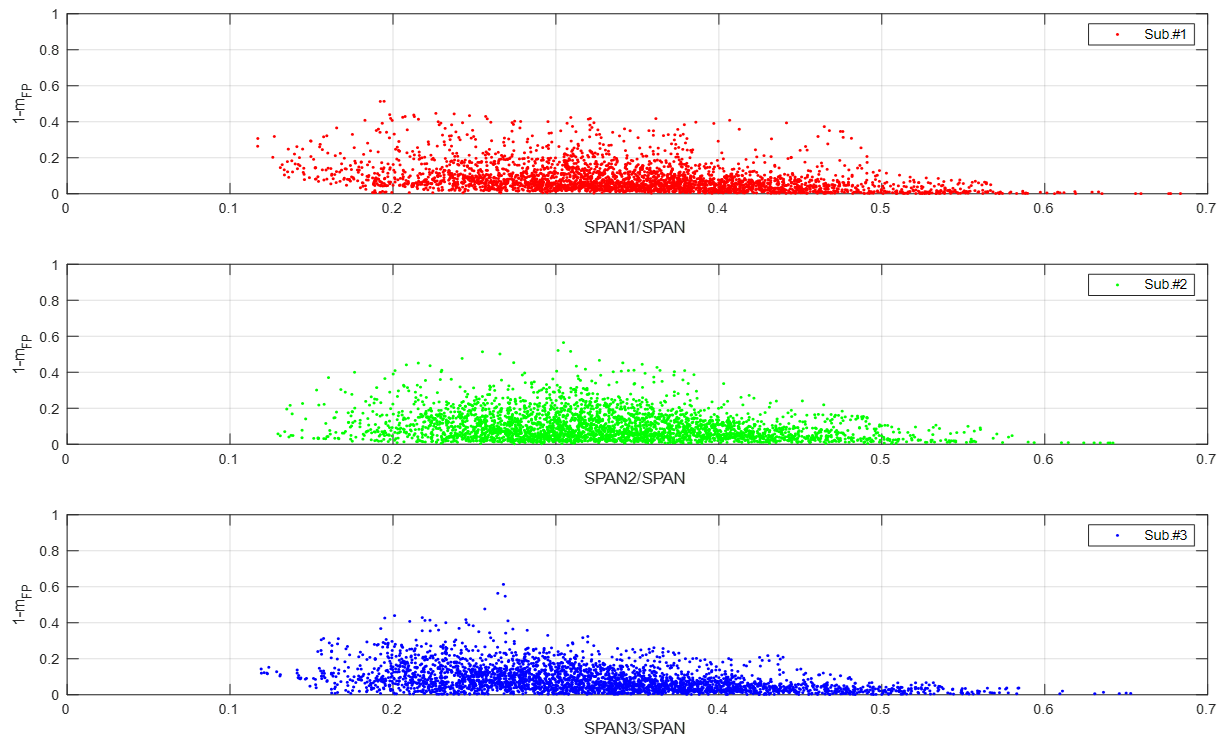}
\end{tabular}
\caption{\small Scatter plots of $1-m_{FP}$ vs. $Span_i/Span$ for all three subapertures (transect G).}
\label{f:scat}
\end{figure*}

In agreement with previous observations, the $1-m_{FP}$ noticeably varies among subapertures. For some parts the full resolution depolarising factor is dominated by one of the subapertures (first and third strong peaks of the transect and around pixels 5200 and 5760, in case of the second subaperture). In other regions there appears a more similar behaviour among all three azimuth look angles or even there could be cases where two out of three subapertures resemble the full resolution pattern whereas the third one is either above or below it. In addition, there also appear some cases where the full resolution $1-m_{FP}$ takes values higher than any of the ones for individual subapertures. On the other hand, the relative contributions of individual subaperture backscattering exhibit a highly variable pattern with high dynamic range. In order to jointly explore the patterns exhibit by both features, the scatter plots of $1-m_{FP}$ vs. $Span_i/Span$ have been represented in Figure \ref{f:scat} for the whole transect G. 
 
In general, these scatter plots exhibit a higher variance of $1-m_{FP}$ at low values of $Span_i/Span$ up to 0.3 or 0.35, i.e. when the relative contribution of the subaperture backscattering to the whole image span is not higher than 35\% . On the contrary, when an individual subaperture collects the highest amount of backscattered power, then the $1-m_{FP}$ factor greatly decreases to values below 0.2 which is compatible with strong polarised radar returns. These scatter plots reveal an average inverse relationship between $1-m_{FP}$ and $Span_i/Span$ which is consistent with potential target structural variations (ignoring terrain slope effects as it is the case) causing anisotropic signatures in vegetation radar scattering.

\begin{figure*}[h!]
 \centering
\begin{tabular}{c}
  \includegraphics[width=16cm]{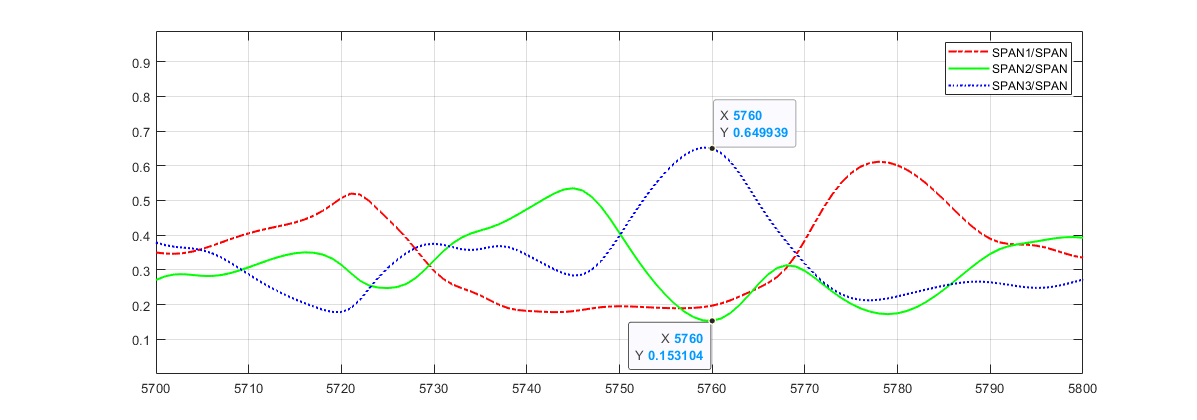}\\
	\includegraphics[width=16cm]{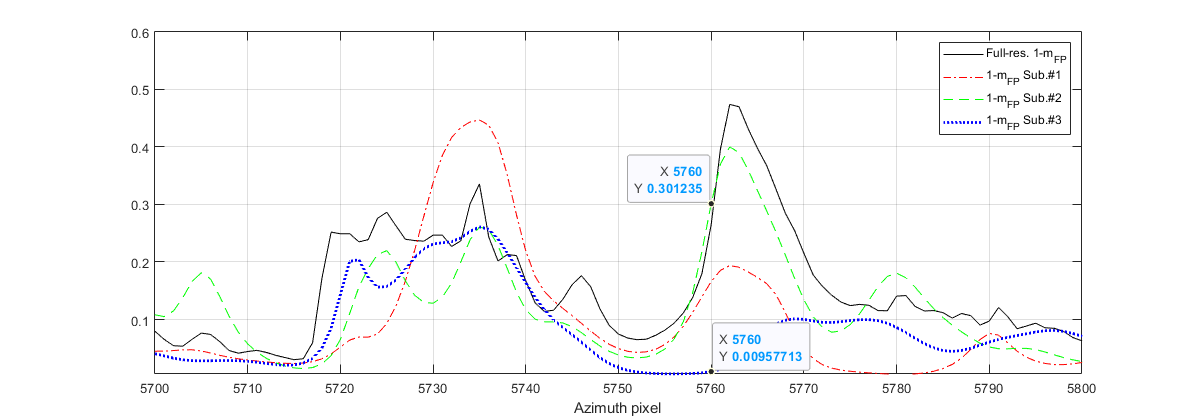}
\end{tabular}
\caption{\small Interval of transect G: (top image) variation of $Span_i/Span$; (bottom image) $1-m_{FP}$ value for full resolution (black solid line) and individual subapertures (different colored line styles).}
\label{f:cas}
\end{figure*}

A particular case to exemplify this behaviour is displayed in Figure \ref{f:cas} where pixel 5760 is considered. Subaperture \#3 gathers 65\% of total span but it is a totally polarised signal, as inferred by the $1-m_{FP}$ value which is less than 0.01. Indeed, the full resolution MF4C decomposition for that pixel yields percentages of 40.6\%, 28.37\%, 26.5\%, and 4.51\% for Pd, Ps, Pv, and Pc, respectively. Hence, the polarised components made up by double-bounce and surface scattering represent around 69\% of the total backscattering. In addition, for the same pixel, it is also pointed out that subaperture \#2 only covers 15.3\% of total span but it senses depolarising effects that mostly determine the depolarising factor $1-m_{FP}$ of the full resolution image, which is 0.265, despite that for subaperture \#1 is as low as 0.16 and even less than 0.01 for subaperture \#3, as mentioned above. This case clearly differs from others also shown in Figure \ref{f:cas}, as for the pixel intervals 5700-5715 and 5730-5740 where the full resolution $1-m_{FP}$ gets closer to the average of all three $1-m_{FP}$ subaperture values. 

\begin{figure*}[h!]
 \centering
\begin{tabular}{c}
  \includegraphics[width=16cm]{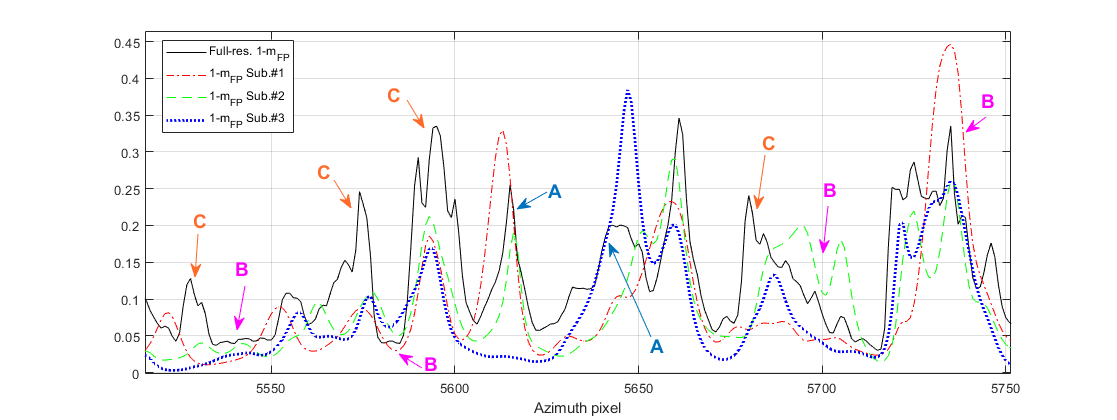}
\end{tabular}
\caption{\small Another interval of transect G illustrating the different situations found when comparing the $1-m_{FP}$ value for full resolution (black solid line) and individual subapertures (different colored line styles). Letters A, B and C represent the following cases regarding the full resolution $1-m_{FP}$, respectively: A) It is dominated by only one individual subaperture; B) It is closer to the average of values from individual subapertures; C) It becomes higher than any individual subaperture value.}
\label{f:cas3}
\end{figure*}

Additionally to the mentioned two situations (i.e. the full resolution $1-m_{FP}$ either dominated by only one individual subaperture or, alternatively, with a value closer to the average of the $1-m_{FP}$ subapertures), there could be also a third case for which the full resolution $1-m_{FP}$ becomes much higher than any of the values for individual subapertures. Figure \ref{f:cas3} displays a portion of transect G where these different behaviours are illustrated. It is noted that letters A, B and C represent, respectively, pixel intervals for cases where $1-m_{FP}$ is dominated by only one individual subaperture, it is closer to the average, or it becomes higher than any individual value.

The conclusion from this particular P-band dataset is twofold. On the one hand, all these observations provide an additional evidence of the anisotropic effects taking place in forest covers at P-band, where the random volume assumption is likely to break down due to local vegetation heterogeneities \cite{ar:garestier08_p,ar:dalessandro2013}. On the other hand,  the reported variation of factor $1-m_{FP}$ for every subaperture in comparison with its value for the full resolution allows us to hypothesise that depolarisation effects even when not being dominant and being only sensed through a small portion of the synthetic aperture could lead to overestimated retrievals of the volume scattering for the full resolution image.

\newpage

\subsection{ALOS PALSAR-1 - Tropical forest}

\begin{figure*}[h!]
 \centering
\begin{tabular}{c}
  \includegraphics[width=12cm]{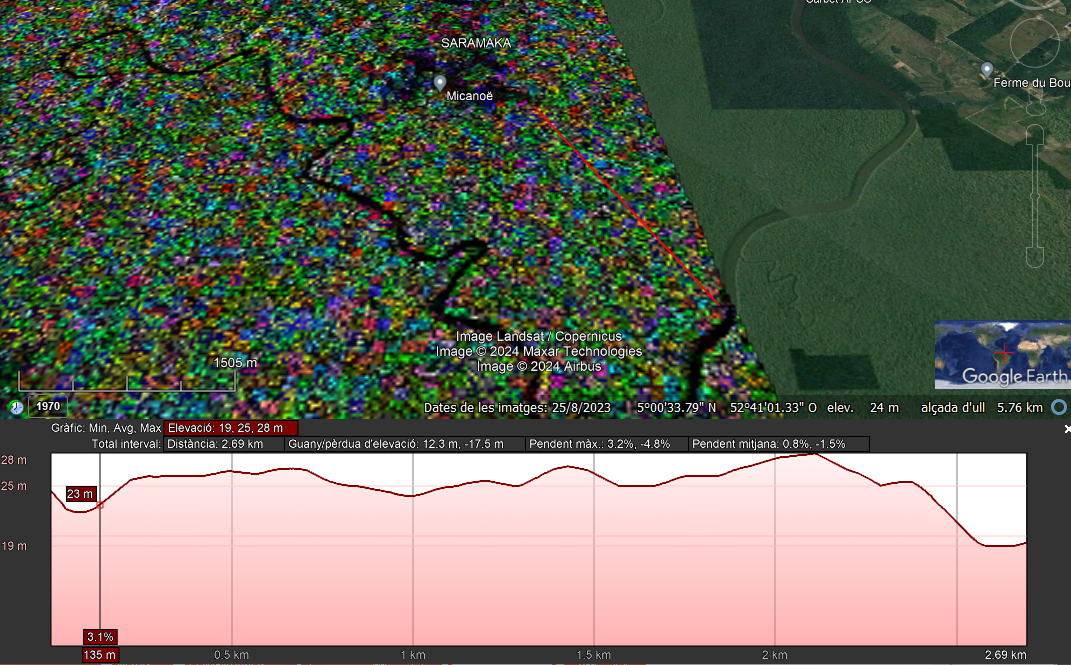}
\end{tabular}
\caption{\small Transect 1 (red line) in Paracou test site. The elevation profile is also given at the bottom of the figure. Maximum and minimum slopes are 3.2\% and -4.8\%, i.e. 1.83$^\circ$ and -2.75$^\circ$, respectively.}
\label{f:para1}
\end{figure*}

Three different transects have been selected for the analysis with this L-band dataset. Figure \ref{f:para1} shows the first one indicated with a red line overlapped on the SAR composition and a Google Earth image. At the bottom of the figure the elevation profile is also plotted. Maximum and minimum values for slopes are 3.2\% and -4.8\%, i.e. 1.83$^\circ$ and -2.75$^\circ$, respectively. Hence, any dominant slope effect potentially coupled in the radar returns can be ruled out. 

\begin{figure*}[h!]
 \centering
\begin{tabular}{c}
  \includegraphics[width=15cm]{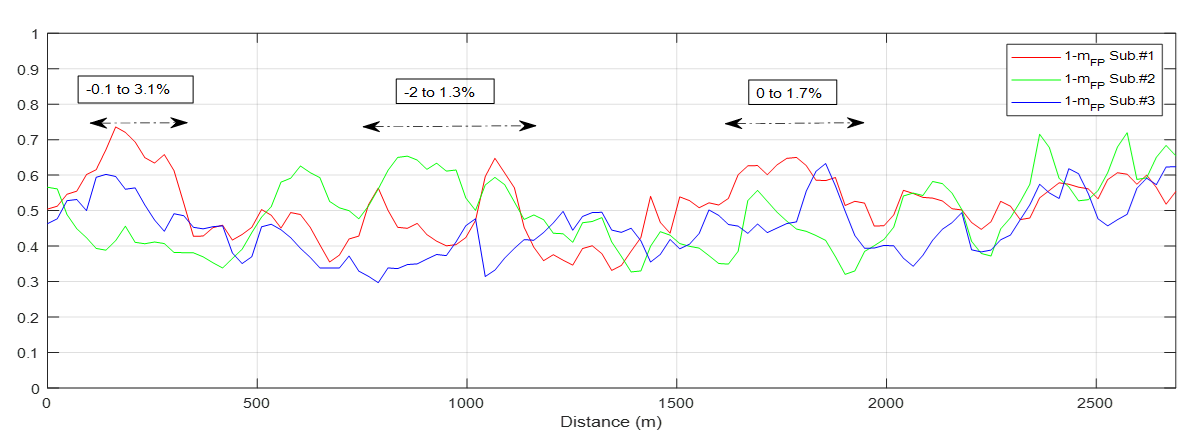}
\end{tabular}
\caption{\small Variation of $1-m_{FP}$ factor for each subaperture for transect 1 in Paracou test site.}
\label{f:para1_prof}
\end{figure*}

The $1-m_{FP}$ factor for every subaperture has been computed and plotted in Figure \ref{f:para1_prof}. In general, it exhibits an increase of $1-m_{FP}$ values due to the stronger vegetation radar signature at L-band in comparison to the previously analysed P-band data, as expected. Some areas exhibiting a high variability among subapertures have been indicated together with the corresponding slopes (in \%). Numerical differences of $1-m_{FP}$ can reach values up to 0.3 which entails a remarkable difference in depolarising effects induced as a function of the azimuth look angle.

\begin{figure*}[h!]
 \centering
\begin{tabular}{c}
  \includegraphics[width=12cm]{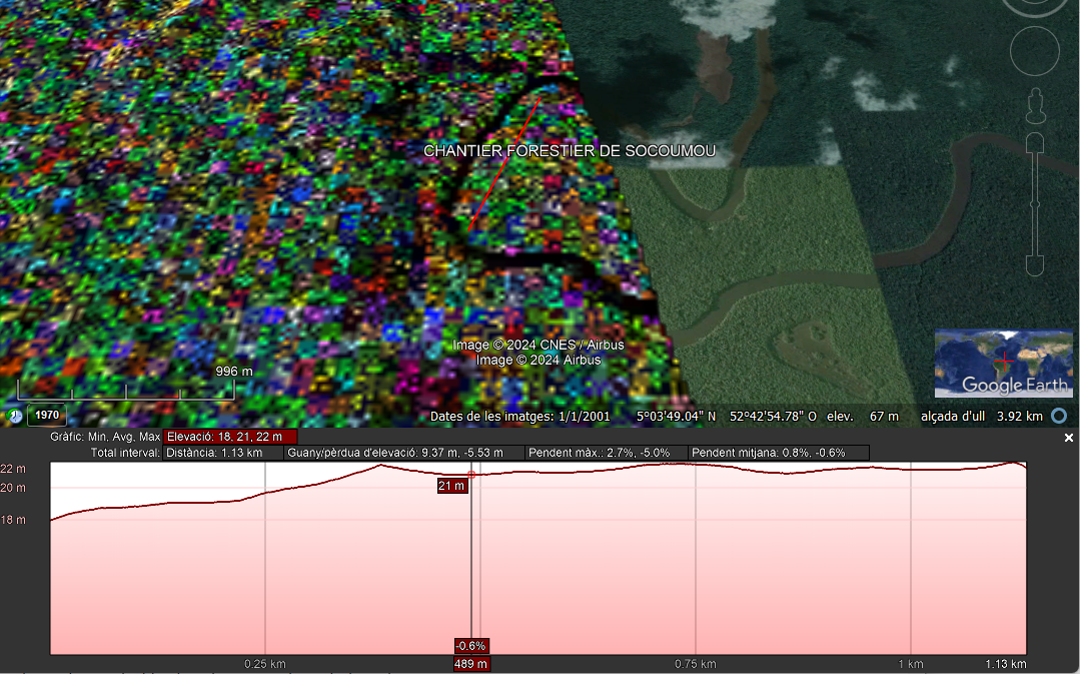}
\end{tabular}
\caption{\small Transect 2 (red line) in Paracou test site. The elevation profile is also given at the bottom of the figure. Maximum and minimum slopes are 2.7\% and -5\%, i.e. 1.54$^\circ$ and -2.86$^\circ$, respectively.}
\label{f:para2}
\end{figure*}

\begin{figure*}[h!]
 \centering
\begin{tabular}{c}
  \includegraphics[width=15cm]{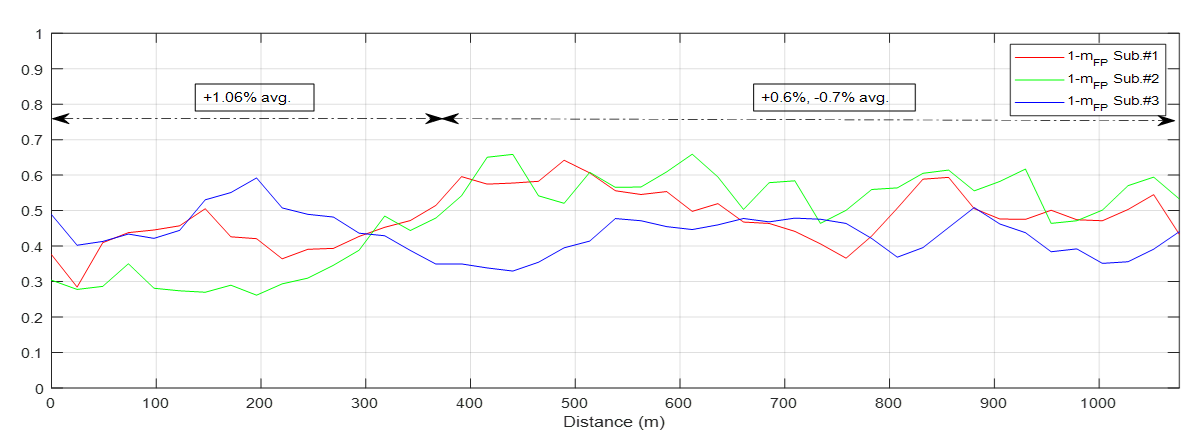}
\end{tabular}
\caption{\small Variation of $1-m_{FP}$ factor for each subaperture for transect 2 in Paracou test site.}
\label{f:para2_prof}
\end{figure*}

Figures \ref{f:para2} and \ref{f:para2_prof} show a second transect where slope values range between 2.7\% and -5\%, i.e. 1.54$^\circ$ and -2.86$^\circ$, respectively. Again, there are some parts where the depolarising effects greatly vary depending on the subaperture with differences over 0.3, in agreement with the outcomes from transect 1.

\begin{figure*}[h!]
 \centering
\begin{tabular}{c}
  \includegraphics[width=16cm]{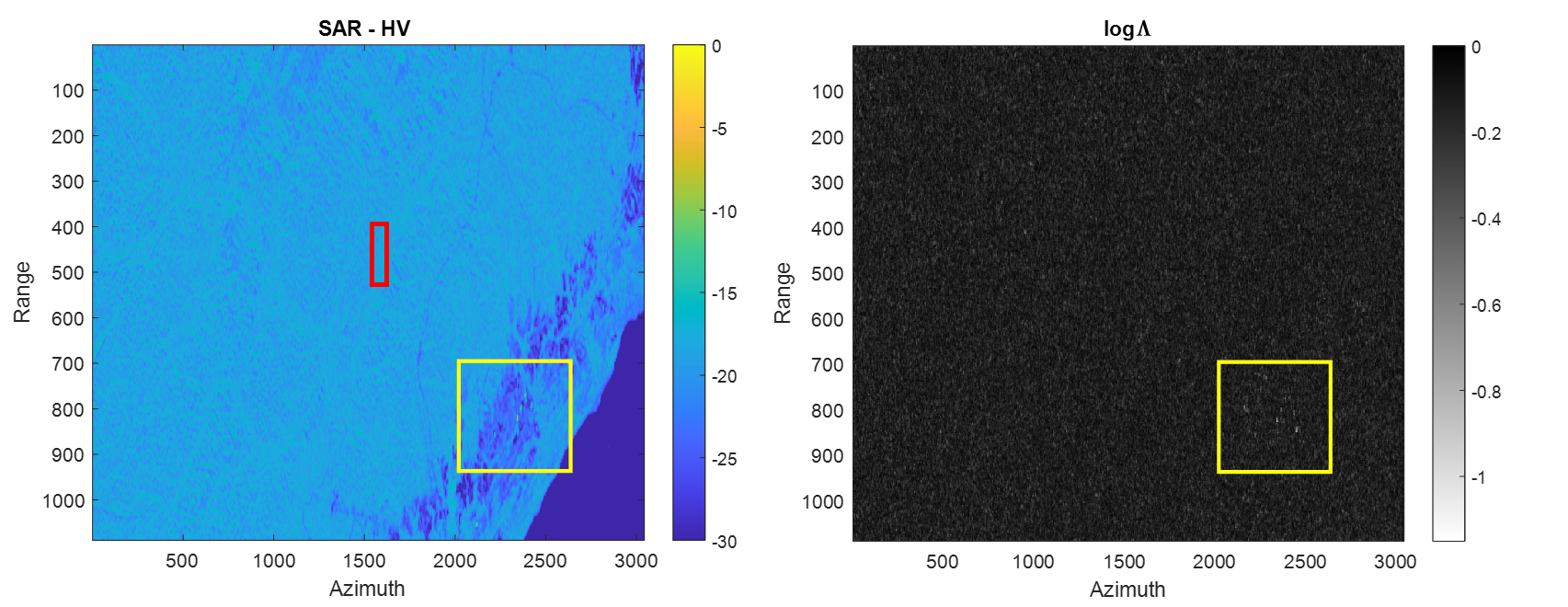}
\end{tabular}
\caption{\small Paracou test site (image appears compressed in azimuth dimension): (Left image) HV channel SAR image shown as a reference; (Right image) Map of the ML ratio ($\log\Lambda$)}
\label{f:lambda_paracou}
\end{figure*}

\begin{figure*}[h!]
 \centering
\begin{tabular}{c}
  \includegraphics[width=8cm]{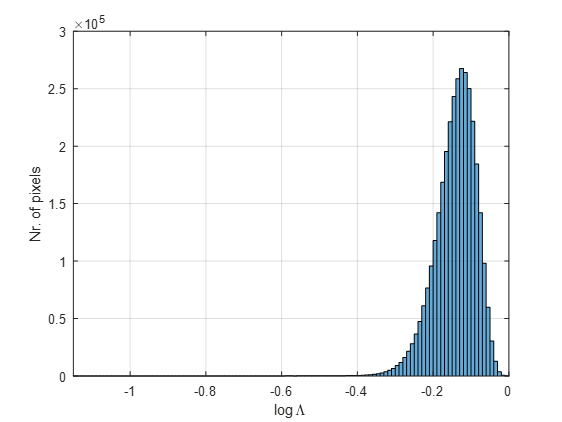}
\end{tabular}
\caption{\small Histogram of $\log\Lambda$ values of the image shown in Figure \ref{f:lambda_paracou}.}
\label{f:hist_paracou}
\end{figure*}

As in the case of P-band data, we have also computed the $\log\Lambda$ indicator which is shown in Figure \ref{f:lambda_paracou} together with a HV channel image for reference. Black level and dark grey shades represent those pixels where the stationarity hypothesis is fulfilled. Expectedly \cite{ar:ferro-famil2003}, this is the case for most of pixels in this image which mostly includes natural covers. The values of $\log\Lambda$ are within an interval ranging from -1.15 to 0, being mostly of them within the [-0.3,0] interval, as shown by the histogram in Figure \ref{f:hist_paracou}. Both the absolute values and dynamic margin of $\log\Lambda$ are noticeably reduced with respect the P-band case. There are some isolated white pixels located inside the yellow rectangle which belong to areas with man-made oriented structures, thus indicating that anisotropic effects dominate there. The main conclusion to be drawn from the $\log\Lambda$ map for this L-band dataset points towards an unquestionable validity of the stationarity assumption and, hence, the absence of orientation effects in the forest cover with regard L-band data. Nevertheless, this conclusion seems to contradict the high variation found at different azimuth observation angles in the depolarising factor $1-m_{FP}$, which can be explained by structural variations of vegetation, in line with our previous analysis at P-band.

\begin{figure*}[h!]
 \centering
\begin{tabular}{c}
  \includegraphics[width=10cm]{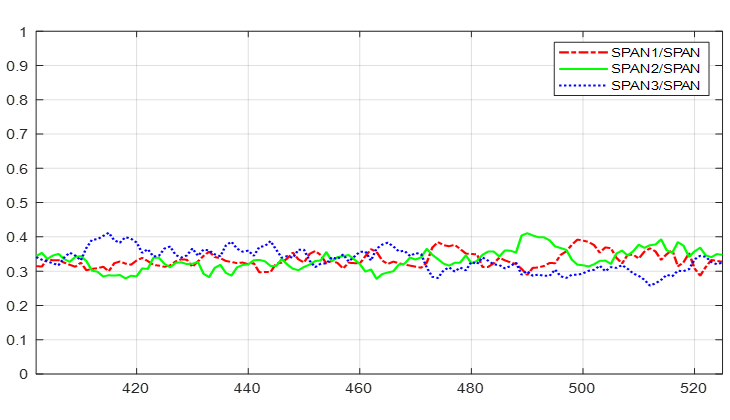}\\
	\includegraphics[width=10cm]{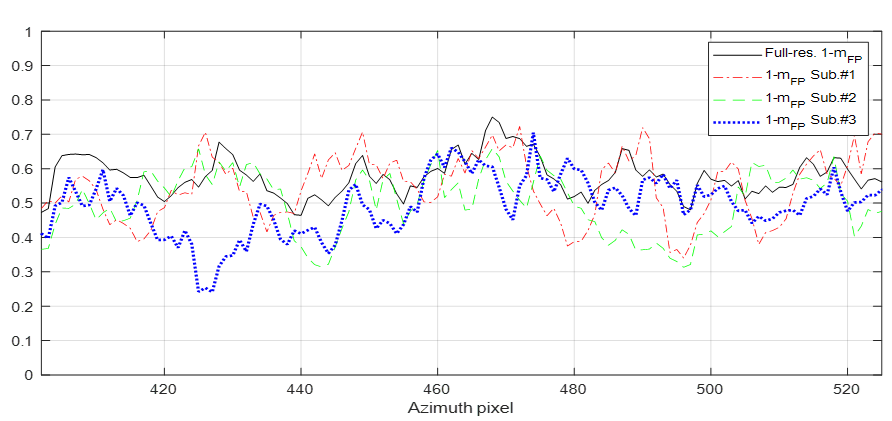}
\end{tabular}
\caption{\small Transect along range dimension indicated in Figure \ref{f:lambda_paracou} with a red rectangle: (Top image) variation of $Span_i/Span$; (Bottom image) $1-m_{FP}$ value for full resolution (black solid line) and individual subapertures (different colored line styles).}
\label{f:transect3}
\end{figure*}

A third transect has been also selected along range covering a lower slope terrain area  within the interval $\pm$2.6\%, i.e. $\pm$1.5$^\circ$. This transect has been indicated with a red rectangle in Figure \ref{f:lambda_paracou}. In this case we have also included in the analysis the $Span_i/Span$ signatures together with the $1-m_{FP}$ variation for all subapertures as well as for the full resolution data. These are displayed in Figure \ref{f:transect3}. 

On the one hand, the $Span_i/Span$ ratios take values quite close to $Span/3$ which represents a proportional distribution of the total span among all three subapertures. This is in agreement with our previous observations on indoor data from vegetation samples (Section \ref{s:indoor}) where from L- and S-band up to X-band the $Span_i/Span$ ratios were almost equal. This fact, however, is in contrast with the clear variation of $Span_i/Span$ along the azimuth look angle shown at P-band (Section \ref{s:biosar}).

On the other hand, the qualitative behaviour exhibited by $1-m_{FP}$ differences among subapertures in Figure \ref{f:transect3} matches those observed in all cases previously presented not only in this section at L-band but also for the two previously analysed datasets. The differences reach values higher than 0.3 which suggests important depolarising effects as a function of the azimuth look angle. In addition, the full resolution $1-m_{FP}$ factor shows the three different behaviours reported above at P-band (see Figure \ref{f:cas3}), i.e. 1) values dominated by only one individual subaperture, 2) values closer to the average, or 3) values higher than any individual value. The overall conclusion from this dataset is not only consistent with findings reported in previous studies \cite{ar:bondur2019} regarding boreal forest heterogeneities at L-band, but it also gives support to the hypothesis that non-dominant depolarisation effects could lead to overestimated volume backscattering power in the full resolution image.

%

%
\section{Discussion}
\label{s:disc}

\subsection{Correction of the volume backscattering power}

The previous analysis based on the subaperture decomposition provides an additional insight for the understanding of the volume backscattering overestimation in PolSAR decompositions. This issue has been object of a intense research (see for example \cite{ar:van_zyl,pro:ainsworth18,ar:ballester_arxiv2020,ar:dey2021,ar:wentao2023}) since the original work by Freeman and Durden \cite{ar:freeman}. Despite the great progress achieved on the topic, no full consensus has yet been reached as to whether there is an optimal way to avoid or minimise this inconsistency, despite the progress made in model-free decompositions \cite{ar:dey2021} or approaches which merge both model-free and model-based schemes \cite{ar:jagdhuber2015,pro:ballester2023}. Therefore, there is still a need for analysing the ranges of applicability of the proposed models \cite{ar:wentao2023}.




Here, we propose a heuristic method for volume component correction based on the assumption that the vegetation anisotropy sensed by any of the subapertures should lead to a decrease of the depolarising effects present in the full resolution image. However, our previous analysis on the $1-m_{FP}$ factor seems to suggest that depolarisation effects only sensed through a small portion of the synthetic aperture and without being a dominant mechanism could lead to a overestimation of the full resolution volume scattering instead. Therefore, we propose to weight every subaperture $1-m_{FP}$ value with the ratio $Span_i/Span$ which corresponds to the proportion of the total backscattered power gathered by any individual subaperture $i$. Hence, the corrected value of the full resolution $1-m_{FP}$ factor is defined as a function of $Span_i/Span$ and the individual $1-m_{FP}^i$ factor as follows:

\begin{equation}
1-m_{FP} = \sum_{i=1}^{Nsub}\frac{Span_i}{Span}\cdot (1-m_{FP}^i)
\label{eq:new_m}
\end{equation}

Once the full resolution $1-m_{FP}$ is recalculated according to Eq.(\ref{eq:new_m}), then the corrected diffuse backscattering power is obtained as in Eq.(\ref{eq:pv}).

\begin{figure*}[h!]
 \centering
\begin{tabular}{c}
  \includegraphics[width=12cm]{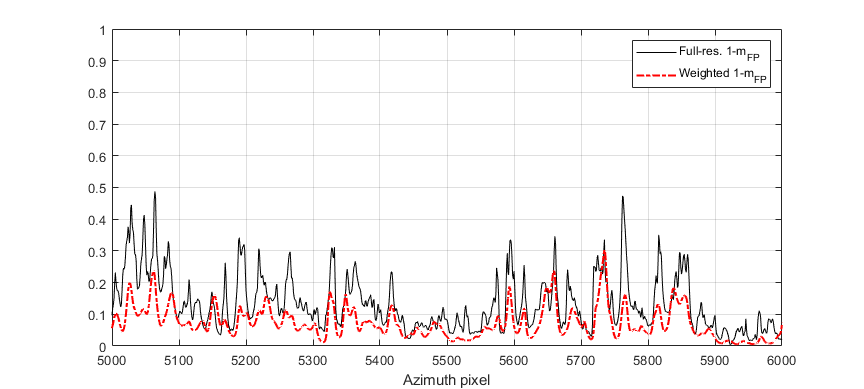}  \\
	\includegraphics[width=12cm]{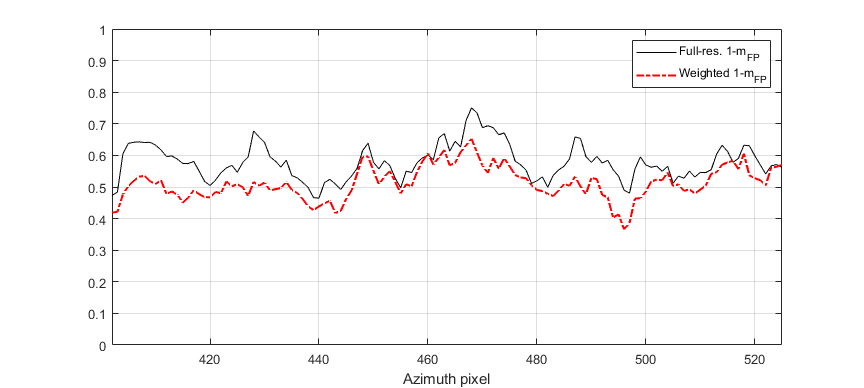}  
\end{tabular}
\caption{\small Values of full resolution $1-m_{FP}$ factor (black solid line) and its corrected version (red dashed line) for (top image) BioSAR2008 data from pixel 5000 to pixel 6000 of transect G, and (bottom image) Paracou test site along transect shown in Figure \ref{f:transect3}.}
\label{f:new_m}
\end{figure*}

We have tested this idea for some transects at P- and L-band previously considered above. Results are shown in Figure \ref{f:new_m}. In both cases, all corrected $1-m_{FP}$ values are lower than the original ones (but for some isolated cases around pixel 5150 or 5700 at P-band, or pixel 460 at L-band). It is observed that differences can reach values up 0.28 at P-band and 0.16 at L-band but, however, there are also some regions in both transects where the corrected $1-m_{FP}$ values do not differ too much from the original ones. This is specially evident at L-band.

According to these new $1-m_{FP}$ values and following the model-free decomposition framework proposed in \cite{ar:dey2021}, this correction leads to the modification of the rest of backscattering powers as they all depend on $m_{FP}$. In particular, it can be seen that one direct consequence of this correction on $1-m_{FP}$ is that the helix backscattering power $P_c$ is inversely modified with respect to the volume backscattering $P_v$. The $P_c$ expression is \cite{ar:dey2021}:

\begin{equation}
P_c = m_{FP} \cdot Span \cdot \sin(2\tau_{FP})
\label{eq:pc}
\end{equation}
where $\tau_{FP}$ is the scattering asymmetry parameter defined in \cite{ar:dey2021}.

As seen in Eq.(\ref{eq:pc}), whenever $1-m_{FP}$ gets lower then $m_{FP}$ is increased and so does the helix component. An expression for the new value of $P_c$, i.e. $P_c^{'}$, in terms of the variation of volume scattering power, $\Delta P_v$, can be obtained as follows. The new volume backscattering $P_v^{'}$ (due to the new depolarising factor) can be defined as:

\begin{equation}
P_v^{'} = P_v + \Delta P_v
\end{equation}

Then, considering the new degree of polarisation $m_{FP}^{'}$ and Eq.(\ref{eq:pv}), the new volume component is:

\begin{equation}
P_v + \Delta P_v = (1 - m_{FP}^{'})\cdot Span
\label{eq:pv2}
\end{equation}

From Eq.(\ref{eq:pv2}), the new degree of polarisation is given as:

\begin{equation}
m_{FP}^{'} = \frac{Span - P_v - \Delta P_v}{Span}
\label{eq:pv3}
\end{equation}

Therefore, by using Eq.(\ref{eq:pc}) and substituting the new expression for the degree of polarisation $m_{FP}^{'}$ therein, and considering that

\begin{equation}
m_{FP} = \frac{Span - P_v}{Span}
\label{eq:pc3}
\end{equation}

and

\begin{equation}
\sin(2\tau_{FP}) = \frac{P_c}{m_{FP} \cdot Span}
\label{eq:pc4}
\end{equation}

then, the new value for the helix component as a function of the variation of volume scattering power, $\Delta P_v$, is written as follows:

\begin{equation}
P_c^{'} = P_c \cdot \left( 1 - \frac{\Delta P_v}{Span - P_v}\right)
\label{eq:pc2}
\end{equation}

Eq.(\ref{eq:pc2}) yields a higher value for the new helix component whenever the new volume power gets reduced (i.e. a negative $\Delta P_v$) or, contrarily, it would be decreased in case that the volume scattering increases after the correction (i.e. a positive $\Delta P_v$).

It is important to bear in mind that this constitutes a preliminary attempt to derive a method to assess the volume overestimation issue and further research on this topic is beyond the scope of this study. Results provided in Figure \ref{f:new_m} only offer an exploratory perspective of the problem and they must be scrutinized in a more systematic way by undertaking additional analysis on larger vegetated areas.

\subsection{Limitations and open issues}

The present study has several limitations that need to be addressed in further work. Firstly, we have employed the subaperture decomposition by assuming three different subapertures with no overlap between adjacent spectra. Previous research \cite{ar:ferro-famil2003,ar:ferro-famil2005,th:sanjuan,ar:sanjuan2015,ar:marino2015} have extensively analysed these issues and found differences depending on these parameters. As we have only examined one possible situation, caution must be applied when interpreting these results. In addition, the degradation of azimuth resolution and the number of samples employed for averaging (which influences both azimuth and range resolution after speckle filtering) have also an impact in the retrieval of scattering mechanisms \cite{ar:garestier08_p} and should be taken into account. Note that, for example, the upcoming ESA's BIOMASS mission is designed to provide 12.5 m and 25 m in azimuth and slant-range coordinates, respectively. Hence, these issues entail reasonable uncertainty regarding the application and performance of any approach based on subaperture decomposition.

Secondly, despite the new insights provided on the overestimation of the vegetation canopy scattering, this study does not demonstrate whether the proposed method for the correction of the depolarising factor $1-m_{FP}$ yields realistic decomposed backscattering powers. This is a drawback also shared by other existing and widely employed methods based on either mathematical or physical constraints and, certainly, it still constitutes a fundamental open issue in this field.

Thirdly, it must taken into account that the width of the azimuth observation interval is the key element in the detection of anisotropies. As is known, different microwave propagation effects are more clearly identified whenever this interval becomes greater. Both the multi-frequency indoor data and the P-band airborne data are inherently acquired under wide azimuth intervals whereas the L-band dataset was acquired by a satellite-borne sensor, thus reducing the azimuthal angle diversity. Despite that in the present results remarkable quantitative differences in the depolarising effects along the azimuth dimension have been observed in all cases, the interpretation provided for P-band data and its practical implication for P-band spaceborne acquisitions must be still confirmed as the anisotropic effects may be diminished under a narrower azimuth interval.

\section{Conclusions and outlook}
\label{s:conc}

This paper has examined the anisotropic behaviour in vegetation canopies and the implications it may have for information retrieval from polarimetric features. To this aim, we have employed the 3-D Barakat degree of polarisation \cite{ar:barakat}, $m_{FP}$, and the subaperture decomposition technique \cite{ar:ferro-famil2003} to quantify the contribution of the diffuse (i.e. volume) scattering power \cite{ar:dey2021} to the total backscattered power from each subaperture. In case of airborne and satellite data we have focused the analysis on areas where terrain slopes are limited approximately to the $\pm$2$^\circ$ interval in order to ensure that no topographic effects are affecting the radar returns. A joint qualitative analysis of the variation of factor $1-m_{FP}$ as a function of the azimuth look angle together with the ratio $Span_i/Span$ (i.e. the ratio of the backscattered power of every single subaperture to the total power from the whole azimuthal integration) has been carried out. Remarkable variations in the $1-m_{FP}$ factor have been observed for all datasets. In case of P-band data, the $1-m_{FP}$ factor exhibits an approximately inversely proportional relationship with $Span_i/Span$ for a particular subaperture. This means that, for this P-band dataset, the presence of anisotropies in the vegetation cover leads to a strongly polarised scattering which manifests itself in a high $Span_i/Span$ ratio and a low value of $1-m_{FP}$ (i.e. below 0.2) for a particular subaperture. However, despite the depolarising effects are not dominant, reasonably high volume scattering values for the full resolution image are retrieved. We have argued that despite these depolarising effects are only sensed through a small portion of the synthetic aperture, they can lead to overestimated retrievals of the volume scattering considering the whole azimuthal integration.

In case of satellite L-band data, the dynamic range of $Span_i/Span$ is much lower and the total backscattered power is almost equally distributed among subapertures, as expected from an isotropic target. However, the $1-m_{FP}$ factor still exhibits significant variations among subapertures whose differences reach values up to 0.3, hence, pointing again towards the presence of structural variations of vegetation and different depolarising effects induced as a function of the azimuth look angle.

Taken together, these conclusions not only agree with previous findings in the existing literature regarding the effect of vegetation heterogeneities on radar scattering, but they also give support to the idea that non-dominant depolarisation effects could lead to overestimated volume backscattering power in the full resolution image. In order to fully confirmed these observations, additional datasets must analysed with the presented methodology.

Finally, according to all these observations, we have speculated with a heuristic approach to correct the volume backscattering power. It is based on the calculation of the corrected full resolution $1-m_{FP}$ factor from a weighted average of the $1-m_{FP}$ factors from all subapertures being the weighting factors the $Span_i/Span$ ratios. This idea has not been fully analysed nor tested in this study and must be object of additional work. In particular, it can be applied in different existing model-based polarimetric decompositions to check up to what extent the remaining negative power issue can be (or not) minimised. 

%
%

%
%


\vspace{2cm}

\section*{Acknowledgements}
The author thanks the EMSL of the European Commission, the European Space Agency (ESA), the German Aerospace Center (DLR), and the Japan Aerospace Exploration Agency (JAXA) for providing the datasets employed in this work. The SLC ALOS PALSAR-1 image has been pre-processed by the free and open access Sentinel Application Platform (SNAP) developed by ESA. The author appreciates the advice of Changcheng Wang on the P-band dataset and the comments of Victor Cazcarra-Bes on the future direction of this research.
\vspace{2cm}

\bibliographystyle{unsrt_reverse}
\bibliography{mycites2,otherabrv,IEEEabrv}

\begin{thebibliography}{10}

\bibitem{ar:treuhaft_cloude}
Treuhaft, R.~N. and Cloude, S.~R.
\newblock The structure of oriented vegetation from polarimetric
  interferometry.
\newblock {\em IEEE Trans. Geosci. Remote Sensing}, 37(5):2620--2624, September
  1999.

\bibitem{ar:ferro-famil2003}
Ferro-Famil, L., Reigber, A., Pottier, E., and Boerner, W.-M.
\newblock Scene characterization using subaperture polarimetric {SAR} data.
\newblock {\em IEEE Transactions on Geoscience and Remote Sensing},
  41(10):2264--2276, 2003.

\bibitem{pro:arnaud}
Arnaud, A.
\newblock Ship detection by sar interferometry.
\newblock In {\em 1999 IEEE International Geoscience and Remote Sensing
  Symposium}, volume~5, pages 2616--2618, Hamburg, Germany, June 1999.

\bibitem{ar:marino2015}
Marino, A., Sanjuan-Ferrer, M., Hajnsek, I., and Ouchi, K.
\newblock Ship detection with spectral analysis of synthetic aperture radar: A
  comparison of new and well-known algorithms.
\newblock {\em Remote Sensing}, 7, April 2015.

\bibitem{ar:sourys2003}
Souyris, J.-C., Henry, C., and Adragna, F.
\newblock On the use of complex {SAR} image spectral analysis for target
  detection: assessment of polarimetry.
\newblock {\em IEEE Transactions on Geoscience and Remote Sensing},
  41(12):2725--2734, 2003.

\bibitem{ar:zandona06}
Schneider, R.~Z., Papathanassiou, K.~P., Hajnsek, I., and Moreira, A.
\newblock Polarimetric and interferometric characterization of coherent
  scatterers in urban areas.
\newblock 44(4):971--984, April 2006.

\bibitem{ar:garestier08_p}
Garestier, F., Dubois-Fernandez, P.~C., and Champion, I.
\newblock Forest height inversion using high-resolution {P}-{B}and
  {P}ol-{InSAR} data.
\newblock {\em IEEE Transactions on Geoscience and Remote Sensing},
  46(11):59--68, 2008.

\bibitem{ar:frey}
Frey, O. and Meier, E.
\newblock Analyzing tomographic {SAR} data of a forest with respect to
  frequency, polarization, and focusing technique.
\newblock {\em IEEE Transactions on Geoscience and Remote Sensing},
  49(10):3648--3659, 2011.

\bibitem{ar:dalessandro2013}
D’Alessandro, M.~M., Tebaldini, S., and Rocca, F.
\newblock Phenomenology of ground scattering in a tropical forest through
  polarimetric synthetic aperture radar tomography.
\newblock {\em IEEE Transactions on Geoscience and Remote Sensing},
  51(8):4430--4437, 2013.

\bibitem{ar:lopez_rvog_test}
Lopez-Martinez, C. and Alonso-Gonzalez, A.
\newblock Assessment and estimation of the {RVoG} model in polarimetric {SAR}
  interferometry.
\newblock {\em IEEE Transactions on Geoscience and Remote Sensing},
  52(6):3091--3106, 2013.

\bibitem{ar:ballester2015}
Ballester-Berman, J.~D., Vicente-Guijalba, F., and Lopez-Sanchez, J.~M.
\newblock A simple {RV}o{G} test for {P}ol{InSAR} data.
\newblock {\em IEEE Journal of Selected Topics in Applied Earth Observations
  and Remote Sensing}, 8(3):1028--1040, 2015.

\bibitem{ar:tello2018}
Tello, M., Cazcarra-Bes, V., Pardini, M., and Papathanassiou, K.
\newblock Forest structure characterization from {SAR} tomography at {L}-band.
\newblock {\em IEEE J. Sel. Topics Appl. Earth Observat. and Remote Sens.},
  11(10):3402–3414, 2018.

\bibitem{ar:bondur2019}
Bondur, V., Chimitdorzhiev, T., Dmitriev, A., and Dagurov, P.
\newblock Spatial anisotropy assessment of the forest vegetation heterogeneity
  at different azimuth angles of radar polarimetric sensing.
\newblock {\em Izvestiya, Atmospheric and Oceanic Physics}, 55:926--934, 2019.

\bibitem{ar:bondur2008}
Bondur, V. and Chimitdorzhiev, T.
\newblock Texture analysis of radar images of vegetation.
\newblock {\em Izv. Vyssh. Uchebn. Zaved., Geod. Aerofotos’emka}, (5):9--14,
  2008.

\bibitem{ar:treuhaft}
Treuhaft, R.~N., Madsen, S.~N., Moghaddam, M., and {van Zyl}, J.~J.
\newblock Vegetation characteristics and underlying topography from
  interferometric radar.
\newblock {\em Radio Science}, 31(6):1449--1485, November 1996.

\bibitem{ar:treuhaft_siqueira}
Treuhaft, R.~N. and Siqueira, P.~R.
\newblock Vertical structure of vegetated land surfaces from interferometric
  and polarimetric data.
\newblock {\em Radio Science}, 35:141--177, 2000.

\bibitem{ar:zhu2023}
Zhu, J., Xie, Y., Fu, H., Wang, C., Wang, H., Liu, Z., and Xie, Q.
\newblock Digital terrain, surface, and canopy height model generation with
  dual-baseline low-frequency {I}n{SAR} over forest areas.
\newblock {\em Journal of Geodesy}, 97(100), 2023.

\bibitem{ar:barakat}
Barakat, R.
\newblock Degree of polarization and the principal idempotents of the coherency
  matrix.
\newblock {\em Opt. Commun.}, 23(2):147--150, 1977.

\bibitem{ar:dey2021}
Dey, S., Bhattacharya, A., Frery, A., López-Martínez, C., and Rao, Y.
\newblock A model-free four component scattering power decomposition for
  polarimetric {SAR} data.
\newblock {\em IEEE Journal of Selected Topics in Applied Earth Observations
  and Remote Sensing}, 14:3887--3902, 2021.

\bibitem{pro:rvi}
Kim, Y. and van Zyl, J.
\newblock Comparison of forest parameter estimation techniques using {SAR}
  data.
\newblock In {\em IEEE 2001 International Geoscience and Remote Sensing
  Symposium (IGARSS 2001}, volume~3, pages 1395--1397, 2001.

\bibitem{ar:szigarski}
Szigarski, C., Jagdhuber, T., Baur, M., Thiel, C., Parrens, M., Wigneron,
  J.-P., Piles, M., and Entekhabi, D.
\newblock Analysis of the radar vegetation index and potential improvements.
\newblock {\em Remote Sensing}, 10(11), 2018.

\bibitem{ar:cloude_letterpct}
Cloude, S.~R.
\newblock Dual baseline coherence tomography.
\newblock {\em IEEE Geoscience and Remote Sensing Letters}, 4(1):127--131,
  January 2007.

\bibitem{ar:coh_tomog}
Cloude, S.~R.
\newblock Polarisation coherence tomography.
\newblock {\em Radio Science}, April 2006.

\bibitem{b:curlander}
Curlander, J.~C. and McDonough, R.~N.
\newblock {\em Synthetic aperture radar: Systems and signal processing}.
\newblock Wiley-Interscience, 1991.

\bibitem{b:lee}
Lee, J.-S. and Pottier, E.
\newblock {\em Polarimetric radar imaging: From Basics to Applications}.
\newblock CRC/Taylor and Francis, 2009.

\bibitem{th:sanjuan}
Sanjuan-Ferrer, M.
\newblock {\em Detection of Coherent Scatterers in SAR Data: Algorithms and
  Applications}.
\newblock PhD thesis, ETH Zurich, Zurich, Switzerland, 2013.

\bibitem{ar:sanjuan2015}
Sanjuan-Ferrer, M.~J., Hajnsek, I., Papathanassiou, K.~P., and Moreira, A.
\newblock A new detection algorithm for coherent scatterers in {SAR} data.
\newblock {\em IEEE Transactions on Geoscience and Remote Sensing},
  53(11):6293--6307, 2015.

\bibitem{web:emsl}
{EMSL} website:
  {https://visitors-centre.jrc.ec.europa.eu/en/media/virtualtours/take-virtual-tour-european-microwave-signature-laboratory}.
\newblock Accessed: 15 March 2024.

\bibitem{b:mitesis}
Lopez-Sanchez, J.~M.
\newblock {\em Analysis and Estimation of Biophysical Parameters of Vegetation
  by Radar Polarimetry}.
\newblock PhD thesis, Polytechnic University of Valencia (UPV), Valencia,
  Spain, January 2000.

\bibitem{th:lluis}
Sagues, L.
\newblock {\em Surface and volumetric scattering analysis of terrains for
  polarimetric and interferometric SAR applications}.
\newblock PhD thesis, Universitat Polit\`ecnica de Catalunya (UPC), Barcelona,
  Spain, September 2000.

\bibitem{ar:ovog}
Ballester-Berman, J.~D., Lopez-Sanchez, J.~M., and Fortuny-Guasch, J.
\newblock Retrieval of biophysical parameters of agricultural crops using
  polarimetric {SAR} interferometry.
\newblock 43(4):683--694, April 2005.

\bibitem{b:biosar2008}
DLR, FOI, CESBIO, and POLIMI.
\newblock {\em Technical Assistance for the Development of Airborne {SAR} and
  Geophysical Measurements during the Bio{SAR} 2008 Experiment}.
\newblock ESA contract No.: 22052/08/NL/CT, 2009.

\bibitem{b:tropisar09}
Dubois-Fernandez, P. and et~al.
\newblock {\em Technical Assistance For The Development of Airborne {SAR} and
  Geophysical Measurements During the {TROPISAR} 2009 Experiment}.
\newblock 2011.

\bibitem{ar:van_zyl}
van Zyl, J.~J., Arii, M., and Kim, Y.
\newblock Model-based decomposition of polarimetric {SAR} covariance matrices
  constrained for nonnegative eigenvalues.
\newblock {\em IEEE Transactions on Geoscience and Remote Sensing},
  49(9):3452--3459, 2011.

\bibitem{pro:ainsworth18}
Ainsworth, T., Lee, J.-S., and Wang, Y.
\newblock Model-based {P}ol{SAR} decompositions: Virtues and vices.
\newblock In {\em 12th European Conference on Synthetic Aperture Radar
  (EUSAR2018)}, pages 8676--8679, Aachen, Germany, June 2018.

\bibitem{ar:ballester_arxiv2020}
Ballester-Berman, J.~D., Ainsworth, T.~L., and Lopez-Sanchez, J.~M.
\newblock {On The Physical Quantitative Assessment of Model-Based PolSAR
  Decompositions}.
\newblock {\em arXiv}, 2001.05872-eess.SP, 2020.

\bibitem{ar:wentao2023}
Wentao~Han, Haiqiang~Fu, J.~Z. and Xie, Q.
\newblock A type of polarimetric parameter for evaluating the reliability of
  model-based decomposition result and its application.
\newblock {\em International Journal of Digital Earth}, 16(1):2111--2128, 2023.

\bibitem{ar:freeman}
Freeman, A. and Durden, S.~L.
\newblock A three-component scattering model for polarimetric {SAR} data.
\newblock 36(3):963--973, May 1998.

\bibitem{ar:jagdhuber2015}
Jagdhuber, T., Hajnsek, I., and Papathanassiou, K.~P.
\newblock An iterative generalized hybrid decomposition for soil moisture
  retrieval under vegetation cover using fully polarimetric {SAR}.
\newblock {\em IEEE Journal of Selected Topics in Applied Earth Observations
  and Remote Sensing}, 8(8):3911--3922, August 2015.

\bibitem{pro:ballester2023}
Ballester-Berman, J.~D.
\newblock Estimation of the copolar phase difference of scattering mechanisms
  in {P}ol{SAR} data.
\newblock https://www.researchgate.net/profile/J-David-Ballester-Berman, June
  2023.
\newblock Working paper, June 2023.

\bibitem{ar:ferro-famil2005}
Ferro-Famil, L., Reigber, A., and Pottier, E.
\newblock Non-stationary natural media analysis from polarimetric {SAR} data
  using a 2-{D} time-frequency decomposition approach.
\newblock {\em Canadian Journal of Remote Sensing}, 31(1), 2005.

\end{thebibliography}

\end{document}